\documentclass[
reprint,
 amsmath,amssymb,
 aps,
]{revtex4-2}
\usepackage{float}
\usepackage{graphicx}
\usepackage{dcolumn}
\usepackage{bm}
\usepackage{upgreek}
\usepackage{lipsum, babel}

\begin{document}

\preprint{APS/123-QED}

\title{Photonic quantum computing on thin-film lithium niobate: Part I Design of an efficient heralded single photon source co-integrated with superconducting detectors}

\author{A. Sayem}
\affiliation{%
 Nokia Bell Labs, USA\\
}%

\date{\today}

\begin{abstract}
Photonic quantum computers are currently one of the primary candidates for fault-tolerant quantum computation. At the heart of the photonic quantum computation lies the strict requirement for suitable quantum sources e.g. high purity, high brightness single photon sources. To build a practical quantum computer, thousands to millions of such sources are required. In this article, we theoretically propose a unique single-photon source design on a thin-film lithium niobate (TFLN) platform co-integrated with superconducting nanowire single-photon detectors. We show that with a judicial design of single photon source using thin film periodically poled lithium waveguides (PPLN), back-illuminated grating couplers (GCs) and directly bonded or integrated cavity coupled superconducting nanowire single-photon detectors (SNSPDs) can lead to a simple but practical high efficiency heralded single-photon source using the current fabrication technology. Such a device will eliminate the requirement of out coupling of the generated photons and can lead to a fully integrated solution. The proposed design can be useful for fusion-based quantum computation and for multiplexed single photon sources and also for efficient on-chip generation and detection of squeezed light. 
\end{abstract}

\maketitle

\section{Introduction}

For linear optical quantum computation, one of the most demanding requirements is the generation of high efficiency, high purity, and high-speed single photons \cite{knill2001scheme,kok2007linear}. Among many single photon generation techniques, currently, the most well-established ones are, i) pseudo-deterministic single-photon generation using quantum dots (QDs) \cite{somaschi2016near,uppu2020scalable} and ii) probabilistic but heralded single-photon generation using non-linear optical parametric processes such as four-wave mixing using $\chi^{(3)}$ materials such as silicon (Si) \cite{paesani2020near}, silicon nitride (SiN) \cite{lu2019chip}, aluminum gallium arsenide (AlGaAs) \cite{steiner2021ultrabright} and three-wave-mixing using $\chi^{(2)}$ materials such as lithium niobate (LN) \cite{zhao2020high}, aluminum nitride (AlN) \cite{guo2017parametric} etc. Though QD can be used to generate single photons in a deterministic fashion \cite{somaschi2016near,uppu2020scalable}, unfortunately, most QD-based single photon sources suffer from low coupling efficiency which in turn make the source probabilistic \cite{vural2020perspective}. 
Also, scaling up such sources is challenging because of the lack of control over the spectral and spatial distribution of the self-assembled QDs \cite{vural2020perspective} and only a few QDs have been integrated on the same chip up to today \cite{papon2023independent}. QDs also suffer from de-phasing which in turn causes the loss of indistinguishably after generating a finite number of photons \cite{maring2023general}.
\begin{figure*}[!ht]
    \centering
    \includegraphics[width=0.8\textwidth]{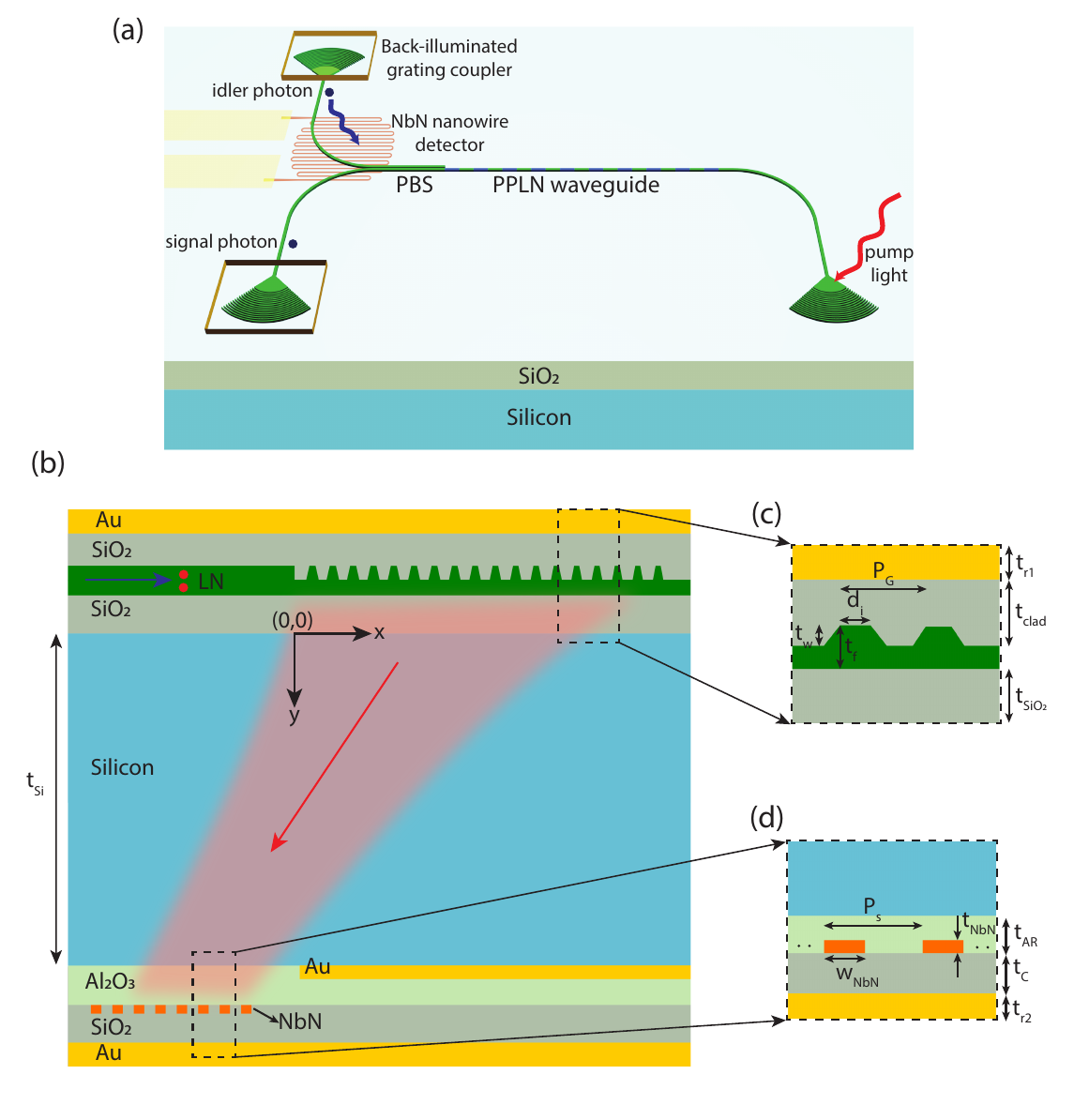}
    \caption{(a) Schematic of the proposed design. A type-II phase-matched PPLN waveguide is coupled to a polarization beam splitter (PBS). One of the generated photons (the idler photon with TE polarization) is guided to an apodized back-illuminated GC cladded with an oxide layer and a top metal reflector. The GC focuses the scattered mode to the backside of the silicon substrate where an NbN-based cavity-coupled SNSPD is either directly fabricated or flip-chip bonded. The signal photon (with different polarization (TM) than the idler photon) is either coupled to a single mode fiber (SMF) through another apodized back-illuminated GC or used for either computation or multiplexing and resource state generation. (b) Cross section of the device structure. (c) Zoomed-in view of the GC structure. The GC has a fixed pitch, $\mathrm{P_{G}}$ with a varying duty cycle, $\mathrm{d_{i}}$ where, i is the number of the grating tooth, the film thickness and etch depth are $\mathrm{t_{f}=600\,nm}$ and $\mathrm{t_{w}=350\,nm}$ respectively. (d) Zoomed-in view of the superconducting cavity detector. The parameters of the detector are provided in table \ref{tab:Tab1} . 
    }
    \label{fig:Fig1}
\end{figure*}
Whereas, though single-photon generation using non-linear optical parametric processes is intrinsically probabilistic in nature, such photon sources always create photons in pairs instead of generating a single photon. This in turn provides the opportunity for heralding e.g. detecting one photon guarantees the presence of the other photon which can be used for computation and hence the loss of the source can be corrected. And with the advancement of mature fabrication technology, it's now possible to fabricate large-scale integrated photonic circuits with a large number of single photon sources \cite{wang2018multidimensional}, and many quantum computing protocols have already been demonstrated \cite{wang2018multidimensional,madsen2022quantum,qiang2018large}. Though tremendous progress has been made on the silicon photonics platform, unfortunately, large-scale fabrication is not the only requirement for fault-tolerant quantum computation. First of all, the propagation loss for silicon photonic circuits is still a concern when compared to that of SiN \cite{liu2021high}, AlN \cite{liu2023aluminum} and even LN \cite{zhang2017monolithic}. Secondly, silicon does not have any Pockels non-linearity, hence thermal or plasma-dispersion effects are typically used to make active devices that are either slow or lossy \cite{rahim2021taking} and most importantly thermal tuning is extremely weak at cryogenic temperature. For photonic quantum computation, high-speed low loss photonic circuits with active tunability is a must not only for multiplexed single-photon sources but also for the implementation of the quantum algorithms and the performing the measurements \cite{bartolucci2021switch,bartolucci2023fusion}. Single photon sources based on optical parametric non-linear processes typically require a strong pump hence extensive pump filtering is required \cite{wang2020integrated,wang2018multidimensional}, and even the deterministic QD sources require pump filtering. Almost all quantum photonic circuits demonstrated so far have used external pump filters \cite{madsen2022quantum,qiang2018large}. Especially for $\chi^{(3)}$ based single-photon sources based on Si or SiN, the signal and idler photons (i.e. the generated photons) are usually close (e.g. hundreds of $\mathrm{\sim GHz}$) to the pump frequency and hence narrow band pump filtering is required \cite{wang2018multidimensional,qiang2018large,madsen2022quantum}. A wide variety of on-chip filters have been proposed such as asymmetric Mach Zehnder interferometer (AMZI) \cite{piekarek2017high}, ring resonators \cite{ong2013ultra}, Bragg-reflectors \cite{harris2014integrated} and so on. Unfortunately, none of these on-chip filters are lossless \cite{wang2020integrated,wang2018multidimensional} and usually require external filtering or multiple cascaded photonic chips \cite{piekarek2017high,harris2014integrated}. The major problem with any external filtering is not primarily the loss associated with the filter but to couple photons in and out of the photonic source chip to the filters and from the filters to the photonic circuit for computation \cite{wang2023deterministic}. As a result, integrating single-photon sources, pump filters, and single-photon detectors on the same chip has been a long-term goal, especially for integrated quantum photonic processing \cite{slussarenko2019photonic}. But due to the extreme demand for pump photon filtering and the complexity of fabrication, integration of single-photon sources, filters, and detectors on the same chip has remained elusive. Instead of using $\chi^{(3)}$ materials such as Si or SiN, $\chi^{(2)}$ materials such as LN \cite{zhu2021integrated}, AlN \cite{liu2023aluminum} are naturally more appealing for two primary reasons. Firstly, it's much easier to filter the pump photons as they are typically one octave higher than the generated photon pairs \cite{zhao2020high,guo2017parametric,xin2022spectrally}. Secondly, they offer Pockels non-linearity which works at any temperature even at cryogenic temperature \cite{lomonte2021single,eltes2020integrated} which opens up the possibility of cryogenic operation of the integrated single photon sources and detectors on the same chip. Among the $\chi^{(2)}$ materials, thin-film lithium niobate (TFLN) is currently the most promising one and has already found many classical applications such as high-speed optical transceivers \cite{wang2018integrated,xu2020high,zhang2021integrated}, frequency comb (FC) generation \cite{zhang2021integrated,wang2019monolithic}, second harmonic (SH) generation \cite{wang2018ultrahigh,lu2021ultralow,sayem2021efficient}, and so on. For quantum optics, lithium niobate is not a new medium \cite{kashiwazaki2020continuous} and with the emergence of TFLN, key ingredients required for quantum computation and communication such as high purity single-photon source \cite{xin2022spectrally}, high-efficiency single photon source \cite{zhao2020high,ma2020ultrabright}, squeezed light generation \cite{chen2022ultra,nehra2022few}, superconducting detector integration and cryogenic reconfigurability \cite{sayem2020lithium,lomonte2021single} has been already demonstrated. One may assume that as single photons sources based on PPLN waveguides and waveguide integrated single photon sources on TFLN have already been successfully demonstrated why not combine those on the same chip and use on-chip filtering \cite{guo201670,piekarek2017high}. Unfortunately, such on-chip filtering doesn't guarantee that the scattered photons won't reach the superconducting nanowires and these scattered photons are typically the reason why cascaded chips are required for high extinction pump filtering \cite{piekarek2017high,harris2014integrated}. And if the photons are coupled out of the photonic chips, we again suffer from the inefficient coupling between the fiber-chip interface. 

In this article, we look for a different approach to integrate single photon sources and detectors. Our goal is to propose a design possible with the current fabrication technology. Taking advantage of the TFLN platform, we theoretically propose and numerically demonstrate a unique design of a heralded single photon source based on thin-film PPLN waveguides, back-illuminated grating couplers \cite{Sayemthesis} and superconducting nanowire single-photon detectors. In our design, the PPLN waveguide is connected to an apodized GC. Using a top metal reflector with a proper oxide cladding on the apodized GC, we show that both pump photons at near visible wavelength (780\,nm) and generated photons at telecom wavelengths (1560\,nm) from the PPLN waveguide can be efficiently guided through the thick ($\mathrm{400\upmu m}$) silicon substrate. The pump photons which are well below the optical bandgap ($\mathrm{Eg_{Si}=1.1\,eV}$) of silicon are fully absorbed by the silicon substrate whereas, the generated photons at the telecom wavelength are guided with negligible loss and focused at the backside of the silicon substrate. Hence, the silicon substrate can act as almost a lossless filter for the single-photon source. At the backside of the silicon substrate, a conventional 1D cavity coupled niobium nitride (NbN) based superconducting nanowire detector is formed which can be either directly fabricated at the backside of the silicon substrate or flip-chip bonded either by void-free Si-Si bonding \cite{takagi1996surface} or by alumina-alumina bonding \cite{takakura2023room,churaev2023heterogeneously}. We show that the combination of the PPLN waveguides, back-illuminated GC, and cavity-coupled superconducting detector at the backside of the substrate silicon can lead to single photon (generated from the PPLN waveguide) detection efficiency as high as $\mathrm{\eta=95\%}$ for the heralding photon. The signal photon can be either used for computation and detected in the same fashion as the idler one or coupled out of the chip efficiently using another back-illuminated GC and a single mode fiber (SMF) \cite{Sayemthesis}. The paper is structured as follows, in section ii), we provide an overall description of the proposed design and describe in detail how the silicon substrate can act as an absorptive filter for the pump photons and calculate the efficiency of such filters, in section iii), we numerically calculate the efficiency of the cavity coupled SNSPD integrated on the backside of the silicon substrate, in section iv), we describe the assumptions made in the simulations and calculation, limitations, and fabrication complexity of the proposed design and probable alternates.

\begin{figure*} [htbp]
    \centering
    \includegraphics[width=1\textwidth]{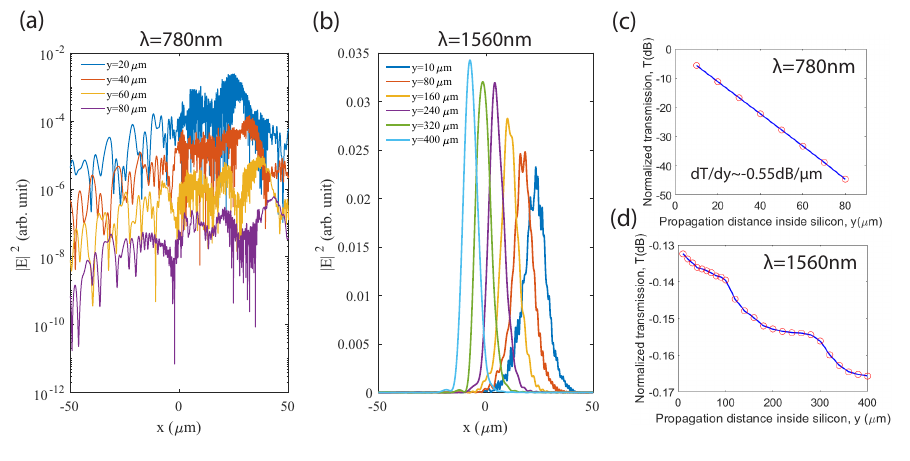}
    \caption{(a) Electric field intensity of the scattered beam from the GC at the signal/idler wavelength, $\mathrm{\lambda=1560\,nm}$ at different propagation distances along the y-axis. (b) Electric field intensity at the pump wavelength, $\mathrm{\lambda=780\,nm}$ at different propagation distances along the y-axis. Normalized optical transmission of the (c) pump and (d) the signal/idler photons as a function of the propagation distance inside the silicon substrate. Here, the transmission is normalized to the power of the input waveguide source at the coupling waveguide as shown by the blue and red arrows in Fig.\ref{fig:Fig2}(a). The GC parameters used in the simulations are, $\mathrm{P_{G}=0.71\,\upmu m}$, $\mathrm{FF_{start}=0.72}$, $\mathrm{FF_{end}=0.2}$, $\mathrm{\alpha=0.012}$, $\mathrm{t_{r1}=200\,nm}$, $\mathrm{t_{clad}=1.02\,\upmu m}$. The total number of periods for the GC is $\mathrm{n=40}$.}
    \label{fig:Fig2}
\end{figure*}

\section{Device Design}

Fig.\ref{fig:Fig1}(a) shows the schematic of the proposed structure. A thin-film PPLN waveguide is followed by a polarization beam splitter (PBS). Here, we assume that the PPLN waveguide is quasi-phase-matched (QPM) for Type-II parametric down-conversion process e.g. the generated photon pairs have different polarization with respect to each other \cite{xin2022spectrally}. In principle, other down-conversion processes such as Type-0 \cite{zhao2020high} or Type-I are also compatible with the proposed design. The reason for choosing the type-II down-conversion process is that single photons with high purity can be generated by using dispersion-engineered PPLN waveguides \cite{xin2022spectrally}. Also, as the generated photons will be of different polarization in each pair, the signal and idler photons can be separated on-chip in a deterministic and efficient way and such high-efficiency PBS has already been successfully demonstrated on the TFLN platform \cite{lin2022high,luo2021high}. In this article, we primarily focus on the rest of the device structure after the PBS. In principle, both X-cut and Z-cut PPLN waveguides will work for the proposed design as long as the pump wavelength, $\lambda_{p}$ is below the optical bandgap of the crystalline silicon substrate. For the type-II down-conversion process considered here, a pump photon at the pump wavelength, $\mathrm{\lambda_{p}=780\,nm}$ is converted to two lower energy photons at $\mathrm{\lambda_{s}=1560\,nm}$ with opposite polarization (i.e. the idler photon with TE polarization and the signal photon with TM polarization). The generated signal and idler photons along with the pump photons are guided by the waveguide and then get split by the PBS. The idler photon and the pump photons are then scattered by the back-illuminated apodized GC as shown in Fig.\ref{fig:Fig1}(a). The GC is designed for the signal/idler photons but not for the pump photons. The GC design consists of apodized gratings along the propagation direction (x-axis) as shown in Fig.\ref{fig:Fig1}(a). The GC is cladded with an oxide (SiO$_{2}$) layer with thickness, $\mathrm{t_{clad}}$ and a top metal (Au) reflector with thickness $\mathrm{t_{r1}}$, to efficiently direct the idler/signal photons in the backward direction towards the silicon substrate. The unit cell of the GC is shown Fig.\ref{fig:Fig1}(c). Each grating tooth has a pitch, $P_{G}$, and a tooth width, $d_{i}$ with a fill factor $FF_{i}$ defined as $FF_{i} = d_{i}/P_{G}$. Here, i represents the grating tooth number. We vary the fill factor linearly with an apodization factor, $\mathrm{\alpha}$ defined as, $\alpha=FF_{\mathrm{i}}-FF_{\mathrm{i+1}}$. We note that in this article, we focus on the detector integration on the backside of the silicon substrate. But the same GC design can be used to efficiently couple light to a single mode fiber (SMF) using an anti-reflection coating on the backside of the silicon substrate. The details of the working principle, design, and experimental results of such back-illuminated GC can be found in chapter V of ref.\cite{Sayemthesis}. Because of the metal reflector on top of the GC, both the pump and the signal/idler photons are guided in the backward direction towards the substrate. Here, at first, we concentrate on the pump photons and their propagation characteristics inside the silicon substrate. The pump photon wavelength, $\mathrm{\lambda_{p}=780\,nm}$ is well below the band-gap of the crystalline silicon substrate, $\mathrm{Eg_{Si}=1.1\,eV}$. As a result, the pump photons will be absorbed by the silicon substrate before reaching the SNSPD. We perform a finite difference time domain (FDTD) simulation in Lumerical to evaluate the filtering efficiency of the silicon substrate. Here, a waveguide source at $\mathrm{\lambda_{p}=780\,nm}$ is used a launch a TE polarized mode into the coupling waveguide as shown in Fig.\ref{fig:Fig1}(a) with a blue arrow. We use power monitors at different depths of the silicon substrate to calculate the attenuation of the pump photons. Fig.\ref{fig:Fig2}(a) shows the intensity distribution of the pump photons at different depths of the silicon substrate. As expected, the scattered pump light doesn't have any exponential or Gaussian shape as the GC is not designed for the pump wavelength. Fig.\ref{fig:Fig2}(c) shows the normalized transmission (i.e. normalized with respect to the incident pump power coupled to the coupling waveguide) of the pump photons as a function of the propagation distance inside the silicon substrate. From the linear fitting of the normalized transmission with respect to the propagation distance, pump attenuation $\mathrm{0.55\,dB/\upmu m}$ is achieved. So $\mathrm{\sim 400\,\upmu m}$ silicon substrate will attenuate the pump by 220\,dB. We note that at cryogenic temperature the optical band gap of crystalline silicon is lower than its value at room temperature \cite{bludau1974temperature}, and hence the absorption coefficient below the band gap energy is also smaller \cite{nguyen2014temperature,noffsinger2012phonon}. 
\begin{figure*}[t!]
    \centering
    \includegraphics[width=1\textwidth]{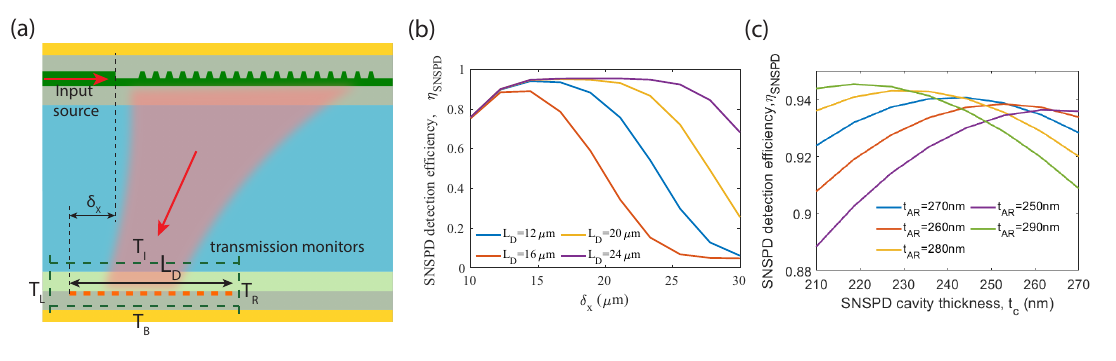}
    \caption{(a) Schematic of the device structure used in FDTD simulation. An input waveguide source as shown by the red arrow is used to launch a waveguide mode with TE polarization to the back-illuminated GC. The mode scattered by the GC is focused on the backside of the silicon substrate. Four transmission monitors were placed on the top, sides, and at bottom of the SNSPD. (b) Absorption of the NbN SNSPD with different lengths, $\mathrm{L_{D}}$ as a function of the detector position along the direction of the GC. The parameters for the cavity SNSPD detector used in the simulation are provided in \ref{tab:Tab1}.}
    \label{fig:Fig3}
\end{figure*}
Even if we assume the absorption coefficient of silicon near the pump wavelength to be half of the value at room temperature \cite{noffsinger2012phonon}, $\mathrm{\sim 400\,\upmu m}$ of silicon substrate will attenuate the pump photons by 110\,dB. Fortunately, the thickness of the substrate silicon of commercially available TFLN is typically within $\mathrm{400\,\upmu m}$ to $\mathrm{600\,\upmu m}$ and in principal, even thicker silicon substrate can be used for the TFLN wafers. From Ref.\cite{xin2022spectrally}, for type-II parametric down-conversion process using thin-film PPLN waveguides, with 220(110)\,dB pump suppression, one would obtain a co-incident to accident count ratio (CAR) of $\mathrm{\sim 3.16X10^{13} (3.16X10^{3})}$. Now, for the signal/idler photons at the telecom wavelength, $\mathrm{\lambda_{p}=1560\,nm}$, there is no attenuation as silicon is transparent at this wavelength \cite{Sayemthesis}. The GC parameters are optimized for maximizing the transmission at the telecom wavelength and we now launch a TE polarized mode using the same waveguide source but at the signal/idler wavelength, $\mathrm{\lambda_{p}=1560\,nm}$. Fig.\ref{fig:Fig2}(b) shows the electric field intensity at the single/idler wavelength, $\mathrm{\lambda_{p}=1560\,nm}$, at different depths of the silicon substrate. Because of the focusing effect due to the apodzied GCs \cite{lomonte2021efficient,Sayemthesis}, the radiated beam form the GC becomes more Gaussian-like as it travels through the silicon substrate. 
In Fig.\ref{fig:Fig2}(d), we plot the normalized transmission (i.e normalized with respect to the indecent signal/idler power coupled to the coupling waveguide) of the signal/idler photons as a function of the propagation distance inside the silicon substrate. From Fig.\ref{fig:Fig2}(d), it can be observed that signal/idler photons suffer less than $\mathrm{\sim0.17\,dB}$ loss while propagating through $\mathrm{\sim 400\,\upmu m}$ thick silicon substrate. Here, significant loss ($\mathrm{\sim0.13\,dB}$) originates directly from the GC  which is primarily due to the back-reflection at the coupling waveguide and GC interface.  

\begin{table*} [!ht]
\caption{\label{tab:Tab1}Parameters used in FDTD simulation for SNSPD efficiency calculation.}
\begin{center}
\begin{tabular}{ c c c c c c c c c c }
\hline
  $t_{AR}$ & $t_{NbN}$ & $P_{S}$ & $w_{NbN}$ & $t_{c}$ & $t_{r2}$ & $(n+ik)_{NbN}$  \\ 
\hline
$275\, nm$ & $5.5\,\mathrm{nm}$ & $200\,\mathrm{nm}$ & $100\,\mathrm{nm}$ & $230\,\mathrm{nm}$ & $200\,\mathrm{nm}$ &  $\mathrm{ 5.23+i5.82}$  \\
\end{tabular}
\end{center}
\end{table*}

\section{Backside integrated cavity coupled SNSPDs}
Fig.\ref{fig:Fig1}(d) shows the unit cell of the cavity-coupled SNSPD. The device structure is very similar to ones proposed before where photons are coupled to a cavity-based superconducting nanowire detector from the substrate side \cite{anant2008optical,rosfjord2006nanowire,tanner2010enhanced,li2014nonideal}. The working principle of the cavity-coupled SNSPDs in Fig.\ref{fig:Fig1}(a) and Fig.\ref{fig:Fig1}(d) is identical to that of a conventional one which has already shown excellent detection efficiency up to 98\,\% \cite{reddy2020superconducting}. Here, in our proposed structure, the primary difference is that photons are not coupled to the cavity detector from an SMF. Rather the photons are generated inside the thin-film PPLN waveguides and then guided to the silicon substrate through an apdodized GC and a top metal reflector. Because of the apodization of the GC, the beam scattered by the GC is focused at the backside of the silicon substrate with a mode field diameter (MFD) similar to that of an SMF \cite{lomonte2021efficient,Sayemthesis}. The cavity-coupled SNSPD consists of an anti-reflection layer with thickness, $\mathrm{t_{AR}}$, a superconducting NbN nanowire with a width, $\mathrm{w_{NbN}}$ and pitch, $\mathrm{P_{s}}$, a cavity layer with thickness, $\mathrm{t_{c}}$ and a metal reflector with thickness $\mathrm{t_{r2}}$. The choice of the materials are $\mathrm{Al_{2}O_{3}}$, $\mathrm{SiO_{2}}$ and $\mathrm{Au}$ for the anti-reflection layer, the cavity and the reflector respectively. Here, we note that in principle, many other material combinations can be used and different deposition techniques such as plasma-enhanced chemical vapor deposition (PECVD), atomic layer deposition (ALD), or sputtering can be used for the deposition of the oxide or the superconducting layers \cite{cheng2019superconducting,esmaeil2021superconducting}. Instead of using any assumption of the radiated mode from the GC, as described in section II, we perform a 2D FDTD simulation in Lumerical with the full device structure including back-illuminated GC, the thick silicon substrate and the cavity coupled SNSPD as shown in Fig.\ref{fig:Fig3}(a). A waveguide source is used to launch a waveguide mode with TE polarization at the signal/idler wavelength at $\mathrm{\lambda_{s,i}=1560\,nm}$ as shown with a red arrow in Fig.\ref{fig:Fig3}(a). The mode is then focused by the apodized GC at the backside of the silicon substrate where the SNSPD is integrated. We calculate the absorption efficiency by the nanowire detector from the transmission monitors located at the top, left, right, and just bottom of the nanowire detectors as shown in Fig.\ref{fig:Fig3}(a). The absorption (i.e. detection) efficiency is then given by, 
\begin{equation}
\eta_{SNSPD}=T_{i}-[T_{L}+T_{R}+T_{B}] 
\label{Eqn1}
\end{equation}
where, $\mathrm{T_{I}}$, $\mathrm{T_{L}}$, $\mathrm{T_{R}}$ and $\mathrm{T_{B}}$ are the transmission coefficients of the power monitors located on the top, left, right and bottom of the nanowire detector normalized to the power of the input waveguide source. The transmission coefficient, $\mathrm{T_{I}}$ includes the transmission loss from the GC as can be seen from Fig.\ref{fig:Fig2}(d). In Fig.\ref{fig:Fig3}(b), we plot the detector efficiency, $\eta_{SNSPD}$ as a function of the detector position along the x-axis with varying total length, $L_{D}$. The other detector parameters such as the pitch, NbN thickness and width, and the cavity parameters are provided in table \ref{tab:Tab1}. As expected, $\eta_{SNSPD}$ reaches the maximum when the detector position coincides with the focused beam radiated by the back-illuminated GC. Also, longer detectors show higher $\eta_{SNSPD}$ as expected but in principle, we would like to use shorter nanowires. Because longer nanowires will have higher kinetic inductance hence the timing response such as the detector speed and timing jitters will be compromised \cite{reddy2020superconducting,natarajan2012superconducting,esmaeil2021superconducting}. Fortunately, as we have shown in section II, we can focus the radiated light at the backside of the substrate. And hence, the detector length, $L_{D}$ required for the saturated efficiency, $\eta_{SNSPD}$ is close to the ones used for fiber-coupled high-efficiency SNSPDs \cite{marsili2013detecting,verma2015high}. In Fig.\ref{fig:Fig3}(c), we plot $\eta_{SNSPD}$ as a function of the SNSPD cavity thickness with varying thicknesses of the anti-reflection coating. From Fig.\ref{fig:Fig3}(c) and Fig.\ref{fig:Fig3}(d), it can be observed that close to $95\,\%$ detection efficiency can be achieved. Here, we assume that the detectors will be well-saturated electrically \cite{gourgues2019superconducting,esmaeil2017single,esmaeil2021superconducting,cheng2019superconducting}.      

\begin{figure*}
    \centering
    \includegraphics[width=1\textwidth]{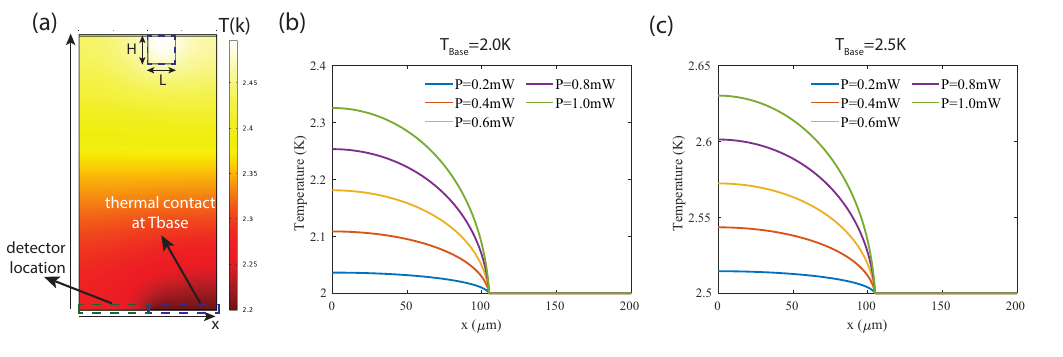}
    \caption{(a) Temperature distribution inside the silicon substrate due to the absorbed pump photons. Here, the base temperature is, $\mathrm{T_{Base}=2.2\,k}$. Temperature distribution at the backside of the silicon substrate at, $\mathrm{y=400\,\upmu m}$ at different base temperature, (b) $\mathrm{T_{Base}=2.0\,k}$ and (c) $\mathrm{T_{Base}=2.5\,k}$}
    \label{fig:Fig4}
\end{figure*}

\section{Assumptions, limitations, and challengers of the proposed design}
In this section, we describe in detail the assumptions used in our simulation, the effect of heat generation due to the absorption of the pump photons, and lastly the fabrication complexities and possible alternates to a fully integrated device. 
\subsection{Assumptions in the simulation}
We first mention the assumptions in our simulations. First of all, we perform the GC transmission and the SNSPD absorption simulation in a two-dimensional (2D) space. The GC simulation is challenging even in a 2D space as it requires fine meshing along with a large simulation space as we need to consider the thick ($\mathrm{400\,\upmu m}$) silicon substrate. But for a GC, 2D simulation is typically enough to correctly estimate the coupling efficiency (i.e. the transmission efficiency) of a GC \cite{lomonte2021efficient, Sayemthesis}. Also, we only consider the focusing of the radiated beam from the GC in the vertical direction (i.e. the y-axis), a linear taper with appropriate length can be used to transversely focus the radiated light from the GC \cite{oton2015long,khan2020long,lomonte2021efficient, Sayemthesis}. 

\subsection{Heat generation due to photon absorption}
\begin{figure*} [!ht]
    \centering
    \includegraphics[width=1\textwidth]{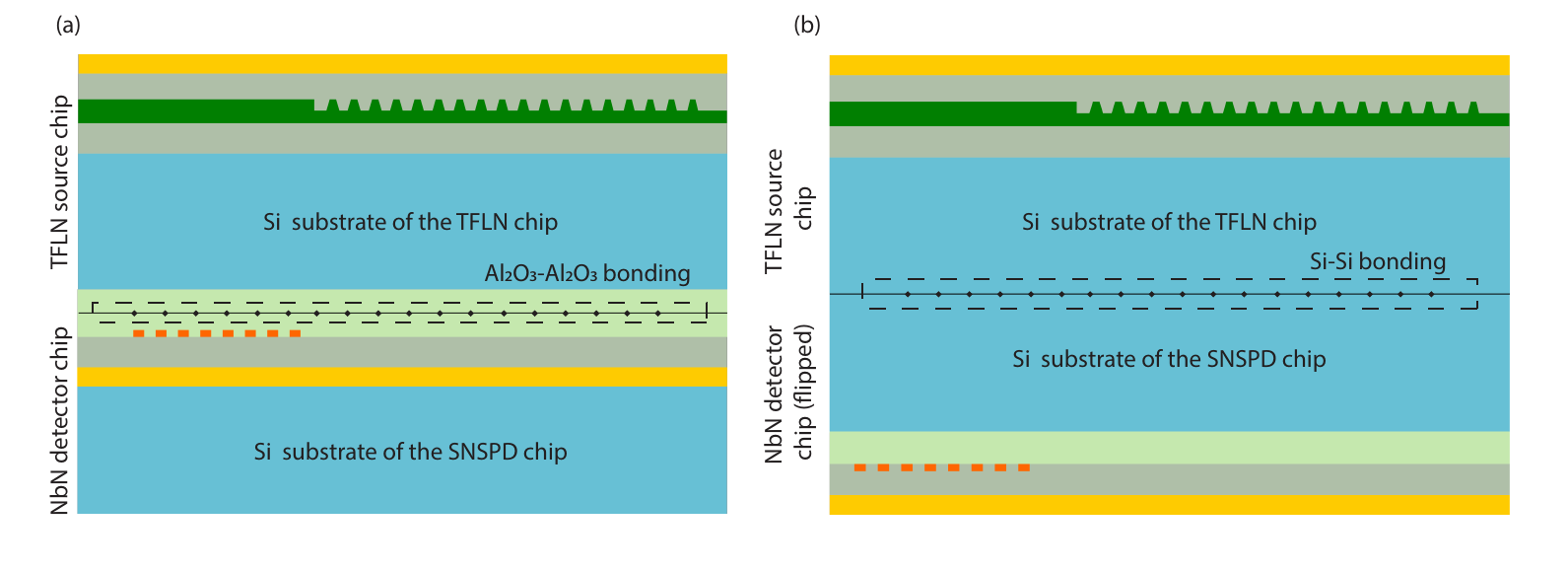}
    \caption{Alternate design for the single photon source-detector integration with (a) $\mathrm{Al_{2}O_{3}}$-$\mathrm{Al_{2}O_{3}}$ bonding} and (b) Si-Si bonding.
    \label{fig:Fig5}
\end{figure*}

The most important thing to consider at first is heat generation due to the absorption of the pump photons (at $\mathrm{\lambda_{p}=780\,nm}$) by the silicon substrate. As the SNSPDs are placed at the backside of the silicon substrate such heat dissipation can increase the operating temperature of the superconducting detectors and hence degrade the performance \cite{gourgues2019superconducting} which can hinder the advantage of the source detector integration. To investigate the effect of heat generation due to the absorption of the pump photons by the silicon substrate, we perform a heat transfer simulation in COMSOL multi-physics using the heat transfer in solids module. From Fig.\ref{fig:Fig3}(c), it can be observed that the pump photons are absorbed at a rate of $\mathrm{0.55dB\,\\upmu m}$ inside the silicon substrate so the $\mathrm{\sim37\,\upmu m}$ will absorb $\mathrm{99\,\%}$ of the pump photons. Hence, here we consider a simplified case for the thermal analysis. We use $\mathrm{H=40\,\upmu m}$ silicon just below the LN GC and the bottom $\mathrm{SiO_{2}}$ layer as the heat source and simulate the temperature distribution as a function of the average pump power (i.e. the pump power absorbed by the silicon substrate). The width of the heat source, L is the same as the size of GC. Here, we assume that the pump power after the PPLN waveguide will be completely absorbed by the silicon substrate which is the upper limit (i.e. the worst-case scenario) of the pump power that will be absorbed by the silicon substrate. Fig.\ref{fig:Fig4}(a) shows the temperature distribution inside the $\mathrm{400\,\upmu m}$ thick silicon substrate, here we assume that the bottom right surface of the chip is thermally anchored to the cold plate of the mixing chamber of the fridge at different base temperatures, $\mathrm{T=2-2.5\,K}$. We also note that there is absolutely no need to use a dilution fridge as mili-Kelvin (mK) temperature is not required for the operation of SNSPDs and high-efficiency detectors ($80\%-90\%\,$) operating as high as 2.5\,K have been demonstrated \cite{verma2015high,gourgues2019superconducting,esmaeil2017single}. For most quantum computation protocols such as measurement-based quantum computation \cite{raussendorf2003measurement,briegel2009measurement} or fusion-based quantum computation \cite{bartolucci2023fusion} and most importantly for multiplexed single-photon sources \cite{bartolucci2021switch} or resource state generators \cite{bartolucci2023fusion}, it is safe to assume that pump will be pulsed instead of continuous \cite{xin2022spectrally,paesani2020near}. We note that high $\mathrm{T_{c}}$ superconducting single-photon detectors have been recently demonstrated experimentally \cite{charaev2023single} but such exotic detectors are not really necessary especially when the goal is to achieve high overall efficiency (photon generation to detection efficiency). In Fig.\ref{fig:Fig5}(b) and Fig.\ref{fig:Fig5}(c) we plot the temperature distribution at the backside of the silicon substrate at, $\mathrm{y=400\,\upmu m}$ at different base temperatures of the cryogenic chamber. We note that during simulation, we used appropriate values for the thermal conductivity of silicon at cryogenic temperature \cite{ziabari2010adaptive}. It can be observed that near the location where the detectors will be integrated, the temperature will increase only by a few hundred of mili-kelvin (mK). As high-efficiency SNSPDs have already been demonstrated experimentally at such high-temperature \cite{verma2015high,gourgues2019superconducting,esmaeil2017single}, we expect such temperature rise won't have any major impact on the performance of the superconducting detectors. Also from Ref.\cite{xin2022spectrally} using type-II down-conversion process and thin-film PPLN waveguides, only $\mathrm{0.33\,mW}$ pump power would be required to generate a photon pair with a probability of 0.1. Using the type-0 down-conversion process, the pump power requirement can be even lower by an order of magnitude \cite{zhang2021integrated}. And needless to mention, some on-chip filtering can also be utilized if necessary. Thermal contact can also be made on the top surface by fully etching LN and the bottom oxide layer around the GC which might be more suitable for thermal cooling which needs to be tested experimentally. Lastly, as non-linear single-photon sources are mostly probabilistic, multiplexing is required to generate a nearly deterministic single-photon source \cite{bartolucci2021switch}. The design proposed in this article can be extended to multiple single photon sources on the same chip with multiple single photon detectors either fabricated or boned to the backside of the chip. One of the major concerns would be then how many such integrated source-detector devices can be put inside a cryogenic chamber. Here, the efficiency of the single photon source comes into play. Because the efficiency determines how much power we can send to the fridge within the cooling power capability of the fridge. Fortunately, cryostats with a cooling capacity of hundreds of milliwatts are now commercially available and we expect the development of cryostats with higher cooling capacity in the near future. 

\subsection{Fabrication complexity and probable alternates}
In this section, we discuss the possible fabrication process flow and alternate approaches to directly fabricating SNSPDs at the backside of the TFLN source chip. First of all, the single-photon source device with backside GC doesn't require the development of any new complex fabrication process other than what has already been demonstrated experimentally \cite{zhao2020high,Sayemthesis}. If we use X-cut TFLN, the first step of the fabrication would be the poling process followed by the PPLN waveguide \cite{zhao2020high,wang2018ultrahigh} and GC etching along with other photonic components such as PBS which requires only one etching of TFLN \cite{lin2022high,luo2021high, Sayemthesis}. For SNSDP integration, there are primarily two ways. The first way would be directly fabricating the detector on the backside of the silicon substrate of the TFLN source chip. Here, we note that TFLN wafers are commercially available on polished high-resistivity silicon substrate so no further processing would be required on the backside of the silicon substrate other than chemical cleaning. Also, chemical mechanical polishing (CMP) can be performed at the backside of the silicon substrate if required. Before detector fabrication, the whole PPLN chip with the GC and the top metal reflector can be first protected by another layer of thick PECVD oxide. As the PPLN waveguide and other photonic components are already cladded with an oxide layer, any additional oxide layer on top of the TFLN device will not affect the device performance such as the operation wavelength of the PPLN waveguide, PBS, and also the back-illuminated GC. For the detector fabrication, deposition of the anti-reflection layer, the cavity layer, and the metal reflector can be done using conventional deposition processes such as ALD, PECVD, and metal liftoff. The crucial part would be the deposition of the superconducting layer such NbN or NbTiN. Fortunately, both ALD and sputtering process have been successfully used to directly integrate NbN and NbTiN superconducting nanowire detectors on TFLN and such deposition has no serious adverse effect on the optical performance of the TFLN photonic devices \cite{sayem2020lithium,lomonte2021single}. So for the proposed design, superconductor deposition isn't a serious concern especially as the deposition will be done on the backside of the silicon substrate. One practical issue would be the characterization of the detectors before completing the full fabrication process. Test GCs can be fabricated without the PPLN waveguides and the PBS to characterize the detectors which is typically done for waveguide-integrated SNSPDs \cite{pernice2012high,sayem2020lithium,lomonte2021single}. 
But instead of directly fabricating the superconducting detectors on the backside of the silicon substrate of the TFLN chip, it's possible to fabricate the source and SNSPD chip separately and then bond two chips (the second way). Such bonding can be performed in two different ways, a) using $\mathrm{Al_{2}O_{3}}$-$\mathrm{Al_{2}O_{3}}$ bonding \cite{takakura2023room,churaev2023heterogeneously} as shown schematically in Fig.\ref{fig:Fig5}(a). The fabrication of the TFLN source chip will be exactly the same as before with one additional step i.e. the deposition of a thin $\mathrm{Al_{2}O_{3}}$ layer at the backside of the TFLN source chip. The superconducting detectors can be fabricated on a separate chip starting a with high resistivity silicon wafer followed by the reflection and the cavity layer deposition, superconducting layer deposition, and patterning, and finally the deposition of the $\mathrm{Al_{2}O_{3}}$ anti-reflection layer \cite{esmaeil2017single,marsili2013detecting}. Such $\mathrm{Al_{2}O_{3}}$-$\mathrm{Al_{2}O_{3}}$ has been successfully used to demonstrate wafer-level bonding between TFLN and SiN photonic circuits \cite{churaev2023heterogeneously}. As the ALD process can be used to deposit the anti-reflection layer (i.e. $\mathrm{Al_{2}O_{3}}$ layer), the thickness can be preciously controlled which is required for high-efficiency detection. Fig.\ref{fig:Fig5}(b) shows the second approach for bonding the source and the detector chip. The fabrication of the TFLN source chip will be exactly the same as for direct integration with no additional steps. The superconducting detectors can be fabricated on a separate chip in the same way but with a slightly different configuration. The detector chip fabrication will start a with high resistivity silicon wafer followed by the anti-reflection layer deposition, superconducting layer deposition, and patterning, the deposition of the cavity layer and the reflection layer \cite{li2014nonideal,rosfjord2006nanowire,tanner2010enhanced,anant2008optical}. Finally, the detector chip can be flipped and bonded to the TFLN source chip by silicon-silicon bonding \cite{amirfeiz2000formation,takagi1996surface}. We note the total silicon substrate thickness doesn't necessarily need to be $\mathrm{400\,\upmu m}$ as we used in this article. The back-illuminated GC can be designed to focus the signal/idler photons at different substrate thicknesses by tuning the GC parameters. Total silicon substrate thickness can be controlled by polishing the silicon substrate of either TFLN source chip or the detector chip or both.  

In conclusion, we propose a new device architecture of a heralded single-photon source on TFLN using PPLN waveguides, back-illuminated GCs, and cavity-coupled superconducting nanowire single-photon detectors. The proposed device can detect generated single photons from the PPLN waveguides with efficiency as high as $\sim95\,\%$. By integrating many such devices on the same chip and using spatial multiplexing and high-speed and low-loss electro-optic active circuits on the TFLN chip, we may envision a near deterministic single photon source and resource state (e.g. Bell states, Greenberger–Horne–Zeilinger states) generation which are the fundamental building blocks of measurement-based photonic quantum computation. We will propose a multiplexing quantum photonic circuit for the generation of a near-deterministic single photon source based on the proposed device in our future work. Also, the proposed design is a direct solution to generate and detect squeezed light on-chip by replacing the single-pixel SNSPD with a multi-pixel detector with photon-number resolving capability \cite{resta2023gigahertz} or by integrating high-efficiency homodyne detectors \cite{Darpa_INSPIRED}. 

\bibliography{main}

\begin{thebibliography}{76}%
\makeatletter
\providecommand \@ifxundefined [1]{%
 \@ifx{#1\undefined}
}%
\providecommand \@ifnum [1]{%
 \ifnum #1\expandafter \@firstoftwo
 \else \expandafter \@secondoftwo
 \fi
}%
\providecommand \@ifx [1]{%
 \ifx #1\expandafter \@firstoftwo
 \else \expandafter \@secondoftwo
 \fi
}%
\providecommand \natexlab [1]{#1}%
\providecommand \enquote  [1]{``#1''}%
\providecommand \bibnamefont  [1]{#1}%
\providecommand \bibfnamefont [1]{#1}%
\providecommand \citenamefont [1]{#1}%
\providecommand \href@noop [0]{\@secondoftwo}%
\providecommand \href [0]{\begingroup \@sanitize@url \@href}%
\providecommand \@href[1]{\@@startlink{#1}\@@href}%
\providecommand \@@href[1]{\endgroup#1\@@endlink}%
\providecommand \@sanitize@url [0]{\catcode `\\12\catcode `\$12\catcode `\&12\catcode `\#12\catcode `\^12\catcode `\_12\catcode `\%12\relax}%
\providecommand \@@startlink[1]{}%
\providecommand \@@endlink[0]{}%
\providecommand \url  [0]{\begingroup\@sanitize@url \@url }%
\providecommand \@url [1]{\endgroup\@href {#1}{\urlprefix }}%
\providecommand \urlprefix  [0]{URL }%
\providecommand \Eprint [0]{\href }%
\providecommand \doibase [0]{https://doi.org/}%
\providecommand \selectlanguage [0]{\@gobble}%
\providecommand \bibinfo  [0]{\@secondoftwo}%
\providecommand \bibfield  [0]{\@secondoftwo}%
\providecommand \translation [1]{[#1]}%
\providecommand \BibitemOpen [0]{}%
\providecommand \bibitemStop [0]{}%
\providecommand \bibitemNoStop [0]{.\EOS\space}%
\providecommand \EOS [0]{\spacefactor3000\relax}%
\providecommand \BibitemShut  [1]{\csname bibitem#1\endcsname}%
\let\auto@bib@innerbib\@empty
\bibitem [{\citenamefont {Knill}\ \emph {et~al.}(2001)\citenamefont {Knill}, \citenamefont {Laflamme},\ and\ \citenamefont {Milburn}}]{knill2001scheme}%
  \BibitemOpen
  \bibfield  {author} {\bibinfo {author} {\bibfnamefont {E.}~\bibnamefont {Knill}}, \bibinfo {author} {\bibfnamefont {R.}~\bibnamefont {Laflamme}},\ and\ \bibinfo {author} {\bibfnamefont {G.~J.}\ \bibnamefont {Milburn}},\ }\bibfield  {title} {\bibinfo {title} {A scheme for efficient quantum computation with linear optics},\ }\href@noop {} {\bibfield  {journal} {\bibinfo  {journal} {nature}\ }\textbf {\bibinfo {volume} {409}},\ \bibinfo {pages} {46} (\bibinfo {year} {2001})}\BibitemShut {NoStop}%
\bibitem [{\citenamefont {Kok}\ \emph {et~al.}(2007)\citenamefont {Kok}, \citenamefont {Munro}, \citenamefont {Nemoto}, \citenamefont {Ralph}, \citenamefont {Dowling},\ and\ \citenamefont {Milburn}}]{kok2007linear}%
  \BibitemOpen
  \bibfield  {author} {\bibinfo {author} {\bibfnamefont {P.}~\bibnamefont {Kok}}, \bibinfo {author} {\bibfnamefont {W.~J.}\ \bibnamefont {Munro}}, \bibinfo {author} {\bibfnamefont {K.}~\bibnamefont {Nemoto}}, \bibinfo {author} {\bibfnamefont {T.~C.}\ \bibnamefont {Ralph}}, \bibinfo {author} {\bibfnamefont {J.~P.}\ \bibnamefont {Dowling}},\ and\ \bibinfo {author} {\bibfnamefont {G.~J.}\ \bibnamefont {Milburn}},\ }\bibfield  {title} {\bibinfo {title} {Linear optical quantum computing with photonic qubits},\ }\href@noop {} {\bibfield  {journal} {\bibinfo  {journal} {Reviews of modern physics}\ }\textbf {\bibinfo {volume} {79}},\ \bibinfo {pages} {135} (\bibinfo {year} {2007})}\BibitemShut {NoStop}%
\bibitem [{\citenamefont {Somaschi}\ \emph {et~al.}(2016)\citenamefont {Somaschi}, \citenamefont {Giesz}, \citenamefont {De~Santis}, \citenamefont {Loredo}, \citenamefont {Almeida}, \citenamefont {Hornecker}, \citenamefont {Portalupi}, \citenamefont {Grange}, \citenamefont {Anton}, \citenamefont {Demory} \emph {et~al.}}]{somaschi2016near}%
  \BibitemOpen
  \bibfield  {author} {\bibinfo {author} {\bibfnamefont {N.}~\bibnamefont {Somaschi}}, \bibinfo {author} {\bibfnamefont {V.}~\bibnamefont {Giesz}}, \bibinfo {author} {\bibfnamefont {L.}~\bibnamefont {De~Santis}}, \bibinfo {author} {\bibfnamefont {J.}~\bibnamefont {Loredo}}, \bibinfo {author} {\bibfnamefont {M.~P.}\ \bibnamefont {Almeida}}, \bibinfo {author} {\bibfnamefont {G.}~\bibnamefont {Hornecker}}, \bibinfo {author} {\bibfnamefont {S.~L.}\ \bibnamefont {Portalupi}}, \bibinfo {author} {\bibfnamefont {T.}~\bibnamefont {Grange}}, \bibinfo {author} {\bibfnamefont {C.}~\bibnamefont {Anton}}, \bibinfo {author} {\bibfnamefont {J.}~\bibnamefont {Demory}}, \emph {et~al.},\ }\bibfield  {title} {\bibinfo {title} {Near-optimal single-photon sources in the solid state},\ }\href@noop {} {\bibfield  {journal} {\bibinfo  {journal} {Nature Photonics}\ }\textbf {\bibinfo {volume} {10}},\ \bibinfo {pages} {340} (\bibinfo {year} {2016})}\BibitemShut {NoStop}%
\bibitem [{\citenamefont {Uppu}\ \emph {et~al.}(2020)\citenamefont {Uppu}, \citenamefont {Pedersen}, \citenamefont {Wang}, \citenamefont {Olesen}, \citenamefont {Papon}, \citenamefont {Zhou}, \citenamefont {Midolo}, \citenamefont {Scholz}, \citenamefont {Wieck}, \citenamefont {Ludwig} \emph {et~al.}}]{uppu2020scalable}%
  \BibitemOpen
  \bibfield  {author} {\bibinfo {author} {\bibfnamefont {R.}~\bibnamefont {Uppu}}, \bibinfo {author} {\bibfnamefont {F.~T.}\ \bibnamefont {Pedersen}}, \bibinfo {author} {\bibfnamefont {Y.}~\bibnamefont {Wang}}, \bibinfo {author} {\bibfnamefont {C.~T.}\ \bibnamefont {Olesen}}, \bibinfo {author} {\bibfnamefont {C.}~\bibnamefont {Papon}}, \bibinfo {author} {\bibfnamefont {X.}~\bibnamefont {Zhou}}, \bibinfo {author} {\bibfnamefont {L.}~\bibnamefont {Midolo}}, \bibinfo {author} {\bibfnamefont {S.}~\bibnamefont {Scholz}}, \bibinfo {author} {\bibfnamefont {A.~D.}\ \bibnamefont {Wieck}}, \bibinfo {author} {\bibfnamefont {A.}~\bibnamefont {Ludwig}}, \emph {et~al.},\ }\bibfield  {title} {\bibinfo {title} {Scalable integrated single-photon source},\ }\href@noop {} {\bibfield  {journal} {\bibinfo  {journal} {Science advances}\ }\textbf {\bibinfo {volume} {6}},\ \bibinfo {pages} {eabc8268} (\bibinfo {year} {2020})}\BibitemShut {NoStop}%
\bibitem [{\citenamefont {Paesani}\ \emph {et~al.}(2020)\citenamefont {Paesani}, \citenamefont {Borghi}, \citenamefont {Signorini}, \citenamefont {Ma{\"\i}nos}, \citenamefont {Pavesi},\ and\ \citenamefont {Laing}}]{paesani2020near}%
  \BibitemOpen
  \bibfield  {author} {\bibinfo {author} {\bibfnamefont {S.}~\bibnamefont {Paesani}}, \bibinfo {author} {\bibfnamefont {M.}~\bibnamefont {Borghi}}, \bibinfo {author} {\bibfnamefont {S.}~\bibnamefont {Signorini}}, \bibinfo {author} {\bibfnamefont {A.}~\bibnamefont {Ma{\"\i}nos}}, \bibinfo {author} {\bibfnamefont {L.}~\bibnamefont {Pavesi}},\ and\ \bibinfo {author} {\bibfnamefont {A.}~\bibnamefont {Laing}},\ }\bibfield  {title} {\bibinfo {title} {Near-ideal spontaneous photon sources in silicon quantum photonics},\ }\href@noop {} {\bibfield  {journal} {\bibinfo  {journal} {Nature communications}\ }\textbf {\bibinfo {volume} {11}},\ \bibinfo {pages} {2505} (\bibinfo {year} {2020})}\BibitemShut {NoStop}%
\bibitem [{\citenamefont {Lu}\ \emph {et~al.}(2019)\citenamefont {Lu}, \citenamefont {Li}, \citenamefont {Westly}, \citenamefont {Moille}, \citenamefont {Singh}, \citenamefont {Anant},\ and\ \citenamefont {Srinivasan}}]{lu2019chip}%
  \BibitemOpen
  \bibfield  {author} {\bibinfo {author} {\bibfnamefont {X.}~\bibnamefont {Lu}}, \bibinfo {author} {\bibfnamefont {Q.}~\bibnamefont {Li}}, \bibinfo {author} {\bibfnamefont {D.~A.}\ \bibnamefont {Westly}}, \bibinfo {author} {\bibfnamefont {G.}~\bibnamefont {Moille}}, \bibinfo {author} {\bibfnamefont {A.}~\bibnamefont {Singh}}, \bibinfo {author} {\bibfnamefont {V.}~\bibnamefont {Anant}},\ and\ \bibinfo {author} {\bibfnamefont {K.}~\bibnamefont {Srinivasan}},\ }\bibfield  {title} {\bibinfo {title} {Chip-integrated visible--telecom entangled photon pair source for quantum communication},\ }\href@noop {} {\bibfield  {journal} {\bibinfo  {journal} {Nature physics}\ }\textbf {\bibinfo {volume} {15}},\ \bibinfo {pages} {373} (\bibinfo {year} {2019})}\BibitemShut {NoStop}%
\bibitem [{\citenamefont {Steiner}\ \emph {et~al.}(2021)\citenamefont {Steiner}, \citenamefont {Castro}, \citenamefont {Chang}, \citenamefont {Dang}, \citenamefont {Xie}, \citenamefont {Norman}, \citenamefont {Bowers},\ and\ \citenamefont {Moody}}]{steiner2021ultrabright}%
  \BibitemOpen
  \bibfield  {author} {\bibinfo {author} {\bibfnamefont {T.~J.}\ \bibnamefont {Steiner}}, \bibinfo {author} {\bibfnamefont {J.~E.}\ \bibnamefont {Castro}}, \bibinfo {author} {\bibfnamefont {L.}~\bibnamefont {Chang}}, \bibinfo {author} {\bibfnamefont {Q.}~\bibnamefont {Dang}}, \bibinfo {author} {\bibfnamefont {W.}~\bibnamefont {Xie}}, \bibinfo {author} {\bibfnamefont {J.}~\bibnamefont {Norman}}, \bibinfo {author} {\bibfnamefont {J.~E.}\ \bibnamefont {Bowers}},\ and\ \bibinfo {author} {\bibfnamefont {G.}~\bibnamefont {Moody}},\ }\bibfield  {title} {\bibinfo {title} {Ultrabright entangled-photon-pair generation from an al ga as-on-insulator microring resonator},\ }\href@noop {} {\bibfield  {journal} {\bibinfo  {journal} {PRX Quantum}\ }\textbf {\bibinfo {volume} {2}},\ \bibinfo {pages} {010337} (\bibinfo {year} {2021})}\BibitemShut {NoStop}%
\bibitem [{\citenamefont {Zhao}\ \emph {et~al.}(2020)\citenamefont {Zhao}, \citenamefont {Ma}, \citenamefont {R{\"u}sing},\ and\ \citenamefont {Mookherjea}}]{zhao2020high}%
  \BibitemOpen
  \bibfield  {author} {\bibinfo {author} {\bibfnamefont {J.}~\bibnamefont {Zhao}}, \bibinfo {author} {\bibfnamefont {C.}~\bibnamefont {Ma}}, \bibinfo {author} {\bibfnamefont {M.}~\bibnamefont {R{\"u}sing}},\ and\ \bibinfo {author} {\bibfnamefont {S.}~\bibnamefont {Mookherjea}},\ }\bibfield  {title} {\bibinfo {title} {High quality entangled photon pair generation in periodically poled thin-film lithium niobate waveguides},\ }\href@noop {} {\bibfield  {journal} {\bibinfo  {journal} {Physical review letters}\ }\textbf {\bibinfo {volume} {124}},\ \bibinfo {pages} {163603} (\bibinfo {year} {2020})}\BibitemShut {NoStop}%
\bibitem [{\citenamefont {Guo}\ \emph {et~al.}(2017)\citenamefont {Guo}, \citenamefont {Zou}, \citenamefont {Schuck}, \citenamefont {Jung}, \citenamefont {Cheng},\ and\ \citenamefont {Tang}}]{guo2017parametric}%
  \BibitemOpen
  \bibfield  {author} {\bibinfo {author} {\bibfnamefont {X.}~\bibnamefont {Guo}}, \bibinfo {author} {\bibfnamefont {C.-l.}\ \bibnamefont {Zou}}, \bibinfo {author} {\bibfnamefont {C.}~\bibnamefont {Schuck}}, \bibinfo {author} {\bibfnamefont {H.}~\bibnamefont {Jung}}, \bibinfo {author} {\bibfnamefont {R.}~\bibnamefont {Cheng}},\ and\ \bibinfo {author} {\bibfnamefont {H.~X.}\ \bibnamefont {Tang}},\ }\bibfield  {title} {\bibinfo {title} {Parametric down-conversion photon-pair source on a nanophotonic chip},\ }\href@noop {} {\bibfield  {journal} {\bibinfo  {journal} {Light: Science \& Applications}\ }\textbf {\bibinfo {volume} {6}},\ \bibinfo {pages} {e16249} (\bibinfo {year} {2017})}\BibitemShut {NoStop}%
\bibitem [{\citenamefont {Vural}\ \emph {et~al.}(2020)\citenamefont {Vural}, \citenamefont {Portalupi},\ and\ \citenamefont {Michler}}]{vural2020perspective}%
  \BibitemOpen
  \bibfield  {author} {\bibinfo {author} {\bibfnamefont {H.}~\bibnamefont {Vural}}, \bibinfo {author} {\bibfnamefont {S.~L.}\ \bibnamefont {Portalupi}},\ and\ \bibinfo {author} {\bibfnamefont {P.}~\bibnamefont {Michler}},\ }\bibfield  {title} {\bibinfo {title} {Perspective of self-assembled ingaas quantum-dots for multi-source quantum implementations},\ }\href@noop {} {\bibfield  {journal} {\bibinfo  {journal} {Applied Physics Letters}\ }\textbf {\bibinfo {volume} {117}} (\bibinfo {year} {2020})}\BibitemShut {NoStop}%
\bibitem [{\citenamefont {Papon}\ \emph {et~al.}(2023)\citenamefont {Papon}, \citenamefont {Wang}, \citenamefont {Uppu}, \citenamefont {Scholz}, \citenamefont {Wieck}, \citenamefont {Ludwig}, \citenamefont {Lodahl},\ and\ \citenamefont {Midolo}}]{papon2023independent}%
  \BibitemOpen
  \bibfield  {author} {\bibinfo {author} {\bibfnamefont {C.}~\bibnamefont {Papon}}, \bibinfo {author} {\bibfnamefont {Y.}~\bibnamefont {Wang}}, \bibinfo {author} {\bibfnamefont {R.}~\bibnamefont {Uppu}}, \bibinfo {author} {\bibfnamefont {S.}~\bibnamefont {Scholz}}, \bibinfo {author} {\bibfnamefont {A.~D.}\ \bibnamefont {Wieck}}, \bibinfo {author} {\bibfnamefont {A.}~\bibnamefont {Ludwig}}, \bibinfo {author} {\bibfnamefont {P.}~\bibnamefont {Lodahl}},\ and\ \bibinfo {author} {\bibfnamefont {L.}~\bibnamefont {Midolo}},\ }\bibfield  {title} {\bibinfo {title} {Independent operation of two waveguide-integrated quantum emitters},\ }\href@noop {} {\bibfield  {journal} {\bibinfo  {journal} {Physical Review Applied}\ }\textbf {\bibinfo {volume} {19}},\ \bibinfo {pages} {L061003} (\bibinfo {year} {2023})}\BibitemShut {NoStop}%
\bibitem [{\citenamefont {Maring}\ \emph {et~al.}(2023)\citenamefont {Maring}, \citenamefont {Fyrillas}, \citenamefont {Pont}, \citenamefont {Ivanov}, \citenamefont {Stepanov}, \citenamefont {Margaria}, \citenamefont {Hease}, \citenamefont {Pishchagin}, \citenamefont {Au}, \citenamefont {Boissier} \emph {et~al.}}]{maring2023general}%
  \BibitemOpen
  \bibfield  {author} {\bibinfo {author} {\bibfnamefont {N.}~\bibnamefont {Maring}}, \bibinfo {author} {\bibfnamefont {A.}~\bibnamefont {Fyrillas}}, \bibinfo {author} {\bibfnamefont {M.}~\bibnamefont {Pont}}, \bibinfo {author} {\bibfnamefont {E.}~\bibnamefont {Ivanov}}, \bibinfo {author} {\bibfnamefont {P.}~\bibnamefont {Stepanov}}, \bibinfo {author} {\bibfnamefont {N.}~\bibnamefont {Margaria}}, \bibinfo {author} {\bibfnamefont {W.}~\bibnamefont {Hease}}, \bibinfo {author} {\bibfnamefont {A.}~\bibnamefont {Pishchagin}}, \bibinfo {author} {\bibfnamefont {T.~H.}\ \bibnamefont {Au}}, \bibinfo {author} {\bibfnamefont {S.}~\bibnamefont {Boissier}}, \emph {et~al.},\ }\bibfield  {title} {\bibinfo {title} {A general-purpose single-photon-based quantum computing platform},\ }\href@noop {} {\bibfield  {journal} {\bibinfo  {journal} {arXiv preprint arXiv:2306.00874}\ } (\bibinfo {year} {2023})}\BibitemShut {NoStop}%
\bibitem [{\citenamefont {Wang}\ \emph {et~al.}(2018{\natexlab{a}})\citenamefont {Wang}, \citenamefont {Paesani}, \citenamefont {Ding}, \citenamefont {Santagati}, \citenamefont {Skrzypczyk}, \citenamefont {Salavrakos}, \citenamefont {Tura}, \citenamefont {Augusiak}, \citenamefont {Man{\v{c}}inska}, \citenamefont {Bacco} \emph {et~al.}}]{wang2018multidimensional}%
  \BibitemOpen
  \bibfield  {author} {\bibinfo {author} {\bibfnamefont {J.}~\bibnamefont {Wang}}, \bibinfo {author} {\bibfnamefont {S.}~\bibnamefont {Paesani}}, \bibinfo {author} {\bibfnamefont {Y.}~\bibnamefont {Ding}}, \bibinfo {author} {\bibfnamefont {R.}~\bibnamefont {Santagati}}, \bibinfo {author} {\bibfnamefont {P.}~\bibnamefont {Skrzypczyk}}, \bibinfo {author} {\bibfnamefont {A.}~\bibnamefont {Salavrakos}}, \bibinfo {author} {\bibfnamefont {J.}~\bibnamefont {Tura}}, \bibinfo {author} {\bibfnamefont {R.}~\bibnamefont {Augusiak}}, \bibinfo {author} {\bibfnamefont {L.}~\bibnamefont {Man{\v{c}}inska}}, \bibinfo {author} {\bibfnamefont {D.}~\bibnamefont {Bacco}}, \emph {et~al.},\ }\bibfield  {title} {\bibinfo {title} {Multidimensional quantum entanglement with large-scale integrated optics},\ }\href@noop {} {\bibfield  {journal} {\bibinfo  {journal} {Science}\ }\textbf {\bibinfo {volume} {360}},\ \bibinfo {pages} {285} (\bibinfo {year} {2018}{\natexlab{a}})}\BibitemShut {NoStop}%
\bibitem [{\citenamefont {Madsen}\ \emph {et~al.}(2022)\citenamefont {Madsen}, \citenamefont {Laudenbach}, \citenamefont {Askarani}, \citenamefont {Rortais}, \citenamefont {Vincent}, \citenamefont {Bulmer}, \citenamefont {Miatto}, \citenamefont {Neuhaus}, \citenamefont {Helt}, \citenamefont {Collins} \emph {et~al.}}]{madsen2022quantum}%
  \BibitemOpen
  \bibfield  {author} {\bibinfo {author} {\bibfnamefont {L.~S.}\ \bibnamefont {Madsen}}, \bibinfo {author} {\bibfnamefont {F.}~\bibnamefont {Laudenbach}}, \bibinfo {author} {\bibfnamefont {M.~F.}\ \bibnamefont {Askarani}}, \bibinfo {author} {\bibfnamefont {F.}~\bibnamefont {Rortais}}, \bibinfo {author} {\bibfnamefont {T.}~\bibnamefont {Vincent}}, \bibinfo {author} {\bibfnamefont {J.~F.}\ \bibnamefont {Bulmer}}, \bibinfo {author} {\bibfnamefont {F.~M.}\ \bibnamefont {Miatto}}, \bibinfo {author} {\bibfnamefont {L.}~\bibnamefont {Neuhaus}}, \bibinfo {author} {\bibfnamefont {L.~G.}\ \bibnamefont {Helt}}, \bibinfo {author} {\bibfnamefont {M.~J.}\ \bibnamefont {Collins}}, \emph {et~al.},\ }\bibfield  {title} {\bibinfo {title} {Quantum computational advantage with a programmable photonic processor},\ }\href@noop {} {\bibfield  {journal} {\bibinfo  {journal} {Nature}\ }\textbf {\bibinfo {volume} {606}},\ \bibinfo {pages} {75} (\bibinfo {year} {2022})}\BibitemShut {NoStop}%
\bibitem [{\citenamefont {Qiang}\ \emph {et~al.}(2018)\citenamefont {Qiang}, \citenamefont {Zhou}, \citenamefont {Wang}, \citenamefont {Wilkes}, \citenamefont {Loke}, \citenamefont {O’Gara}, \citenamefont {Kling}, \citenamefont {Marshall}, \citenamefont {Santagati}, \citenamefont {Ralph} \emph {et~al.}}]{qiang2018large}%
  \BibitemOpen
  \bibfield  {author} {\bibinfo {author} {\bibfnamefont {X.}~\bibnamefont {Qiang}}, \bibinfo {author} {\bibfnamefont {X.}~\bibnamefont {Zhou}}, \bibinfo {author} {\bibfnamefont {J.}~\bibnamefont {Wang}}, \bibinfo {author} {\bibfnamefont {C.~M.}\ \bibnamefont {Wilkes}}, \bibinfo {author} {\bibfnamefont {T.}~\bibnamefont {Loke}}, \bibinfo {author} {\bibfnamefont {S.}~\bibnamefont {O’Gara}}, \bibinfo {author} {\bibfnamefont {L.}~\bibnamefont {Kling}}, \bibinfo {author} {\bibfnamefont {G.~D.}\ \bibnamefont {Marshall}}, \bibinfo {author} {\bibfnamefont {R.}~\bibnamefont {Santagati}}, \bibinfo {author} {\bibfnamefont {T.~C.}\ \bibnamefont {Ralph}}, \emph {et~al.},\ }\bibfield  {title} {\bibinfo {title} {Large-scale silicon quantum photonics implementing arbitrary two-qubit processing},\ }\href@noop {} {\bibfield  {journal} {\bibinfo  {journal} {Nature photonics}\ }\textbf {\bibinfo {volume} {12}},\ \bibinfo {pages} {534} (\bibinfo {year} {2018})}\BibitemShut {NoStop}%
\bibitem [{\citenamefont {Liu}\ \emph {et~al.}(2021)\citenamefont {Liu}, \citenamefont {Huang}, \citenamefont {Wang}, \citenamefont {He}, \citenamefont {Raja}, \citenamefont {Liu}, \citenamefont {Engelsen},\ and\ \citenamefont {Kippenberg}}]{liu2021high}%
  \BibitemOpen
  \bibfield  {author} {\bibinfo {author} {\bibfnamefont {J.}~\bibnamefont {Liu}}, \bibinfo {author} {\bibfnamefont {G.}~\bibnamefont {Huang}}, \bibinfo {author} {\bibfnamefont {R.~N.}\ \bibnamefont {Wang}}, \bibinfo {author} {\bibfnamefont {J.}~\bibnamefont {He}}, \bibinfo {author} {\bibfnamefont {A.~S.}\ \bibnamefont {Raja}}, \bibinfo {author} {\bibfnamefont {T.}~\bibnamefont {Liu}}, \bibinfo {author} {\bibfnamefont {N.~J.}\ \bibnamefont {Engelsen}},\ and\ \bibinfo {author} {\bibfnamefont {T.~J.}\ \bibnamefont {Kippenberg}},\ }\bibfield  {title} {\bibinfo {title} {High-yield, wafer-scale fabrication of ultralow-loss, dispersion-engineered silicon nitride photonic circuits},\ }\href@noop {} {\bibfield  {journal} {\bibinfo  {journal} {Nature communications}\ }\textbf {\bibinfo {volume} {12}},\ \bibinfo {pages} {2236} (\bibinfo {year} {2021})}\BibitemShut {NoStop}%
\bibitem [{\citenamefont {Liu}\ \emph {et~al.}(2023)\citenamefont {Liu}, \citenamefont {Bruch},\ and\ \citenamefont {Tang}}]{liu2023aluminum}%
  \BibitemOpen
  \bibfield  {author} {\bibinfo {author} {\bibfnamefont {X.}~\bibnamefont {Liu}}, \bibinfo {author} {\bibfnamefont {A.~W.}\ \bibnamefont {Bruch}},\ and\ \bibinfo {author} {\bibfnamefont {H.~X.}\ \bibnamefont {Tang}},\ }\bibfield  {title} {\bibinfo {title} {Aluminum nitride photonic integrated circuits: from piezo-optomechanics to nonlinear optics},\ }\href@noop {} {\bibfield  {journal} {\bibinfo  {journal} {Advances in Optics and Photonics}\ }\textbf {\bibinfo {volume} {15}},\ \bibinfo {pages} {236} (\bibinfo {year} {2023})}\BibitemShut {NoStop}%
\bibitem [{\citenamefont {Zhang}\ \emph {et~al.}(2017)\citenamefont {Zhang}, \citenamefont {Wang}, \citenamefont {Cheng}, \citenamefont {Shams-Ansari},\ and\ \citenamefont {Lon{\v{c}}ar}}]{zhang2017monolithic}%
  \BibitemOpen
  \bibfield  {author} {\bibinfo {author} {\bibfnamefont {M.}~\bibnamefont {Zhang}}, \bibinfo {author} {\bibfnamefont {C.}~\bibnamefont {Wang}}, \bibinfo {author} {\bibfnamefont {R.}~\bibnamefont {Cheng}}, \bibinfo {author} {\bibfnamefont {A.}~\bibnamefont {Shams-Ansari}},\ and\ \bibinfo {author} {\bibfnamefont {M.}~\bibnamefont {Lon{\v{c}}ar}},\ }\bibfield  {title} {\bibinfo {title} {Monolithic ultra-high-q lithium niobate microring resonator},\ }\href@noop {} {\bibfield  {journal} {\bibinfo  {journal} {Optica}\ }\textbf {\bibinfo {volume} {4}},\ \bibinfo {pages} {1536} (\bibinfo {year} {2017})}\BibitemShut {NoStop}%
\bibitem [{\citenamefont {Rahim}\ \emph {et~al.}(2021)\citenamefont {Rahim}, \citenamefont {Hermans}, \citenamefont {Wohlfeil}, \citenamefont {Petousi}, \citenamefont {Kuyken}, \citenamefont {Van~Thourhout},\ and\ \citenamefont {Baets}}]{rahim2021taking}%
  \BibitemOpen
  \bibfield  {author} {\bibinfo {author} {\bibfnamefont {A.}~\bibnamefont {Rahim}}, \bibinfo {author} {\bibfnamefont {A.}~\bibnamefont {Hermans}}, \bibinfo {author} {\bibfnamefont {B.}~\bibnamefont {Wohlfeil}}, \bibinfo {author} {\bibfnamefont {D.}~\bibnamefont {Petousi}}, \bibinfo {author} {\bibfnamefont {B.}~\bibnamefont {Kuyken}}, \bibinfo {author} {\bibfnamefont {D.}~\bibnamefont {Van~Thourhout}},\ and\ \bibinfo {author} {\bibfnamefont {R.}~\bibnamefont {Baets}},\ }\bibfield  {title} {\bibinfo {title} {Taking silicon photonics modulators to a higher performance level: state-of-the-art and a review of new technologies},\ }\href@noop {} {\bibfield  {journal} {\bibinfo  {journal} {Advanced Photonics}\ }\textbf {\bibinfo {volume} {3}},\ \bibinfo {pages} {024003} (\bibinfo {year} {2021})}\BibitemShut {NoStop}%
\bibitem [{\citenamefont {Bartolucci}\ \emph {et~al.}(2021)\citenamefont {Bartolucci}, \citenamefont {Birchall}, \citenamefont {Bonneau}, \citenamefont {Cable}, \citenamefont {Gimeno-Segovia}, \citenamefont {Kieling}, \citenamefont {Nickerson}, \citenamefont {Rudolph},\ and\ \citenamefont {Sparrow}}]{bartolucci2021switch}%
  \BibitemOpen
  \bibfield  {author} {\bibinfo {author} {\bibfnamefont {S.}~\bibnamefont {Bartolucci}}, \bibinfo {author} {\bibfnamefont {P.}~\bibnamefont {Birchall}}, \bibinfo {author} {\bibfnamefont {D.}~\bibnamefont {Bonneau}}, \bibinfo {author} {\bibfnamefont {H.}~\bibnamefont {Cable}}, \bibinfo {author} {\bibfnamefont {M.}~\bibnamefont {Gimeno-Segovia}}, \bibinfo {author} {\bibfnamefont {K.}~\bibnamefont {Kieling}}, \bibinfo {author} {\bibfnamefont {N.}~\bibnamefont {Nickerson}}, \bibinfo {author} {\bibfnamefont {T.}~\bibnamefont {Rudolph}},\ and\ \bibinfo {author} {\bibfnamefont {C.}~\bibnamefont {Sparrow}},\ }\bibfield  {title} {\bibinfo {title} {Switch networks for photonic fusion-based quantum computing},\ }\href@noop {} {\bibfield  {journal} {\bibinfo  {journal} {arXiv preprint arXiv:2109.13760}\ } (\bibinfo {year} {2021})}\BibitemShut {NoStop}%
\bibitem [{\citenamefont {Bartolucci}\ \emph {et~al.}(2023)\citenamefont {Bartolucci}, \citenamefont {Birchall}, \citenamefont {Bombin}, \citenamefont {Cable}, \citenamefont {Dawson}, \citenamefont {Gimeno-Segovia}, \citenamefont {Johnston}, \citenamefont {Kieling}, \citenamefont {Nickerson}, \citenamefont {Pant} \emph {et~al.}}]{bartolucci2023fusion}%
  \BibitemOpen
  \bibfield  {author} {\bibinfo {author} {\bibfnamefont {S.}~\bibnamefont {Bartolucci}}, \bibinfo {author} {\bibfnamefont {P.}~\bibnamefont {Birchall}}, \bibinfo {author} {\bibfnamefont {H.}~\bibnamefont {Bombin}}, \bibinfo {author} {\bibfnamefont {H.}~\bibnamefont {Cable}}, \bibinfo {author} {\bibfnamefont {C.}~\bibnamefont {Dawson}}, \bibinfo {author} {\bibfnamefont {M.}~\bibnamefont {Gimeno-Segovia}}, \bibinfo {author} {\bibfnamefont {E.}~\bibnamefont {Johnston}}, \bibinfo {author} {\bibfnamefont {K.}~\bibnamefont {Kieling}}, \bibinfo {author} {\bibfnamefont {N.}~\bibnamefont {Nickerson}}, \bibinfo {author} {\bibfnamefont {M.}~\bibnamefont {Pant}}, \emph {et~al.},\ }\bibfield  {title} {\bibinfo {title} {Fusion-based quantum computation},\ }\href@noop {} {\bibfield  {journal} {\bibinfo  {journal} {Nature Communications}\ }\textbf {\bibinfo {volume} {14}},\ \bibinfo {pages} {912} (\bibinfo {year} {2023})}\BibitemShut {NoStop}%
\bibitem [{\citenamefont {Wang}\ \emph {et~al.}(2020)\citenamefont {Wang}, \citenamefont {Sciarrino}, \citenamefont {Laing},\ and\ \citenamefont {Thompson}}]{wang2020integrated}%
  \BibitemOpen
  \bibfield  {author} {\bibinfo {author} {\bibfnamefont {J.}~\bibnamefont {Wang}}, \bibinfo {author} {\bibfnamefont {F.}~\bibnamefont {Sciarrino}}, \bibinfo {author} {\bibfnamefont {A.}~\bibnamefont {Laing}},\ and\ \bibinfo {author} {\bibfnamefont {M.~G.}\ \bibnamefont {Thompson}},\ }\bibfield  {title} {\bibinfo {title} {Integrated photonic quantum technologies},\ }\href@noop {} {\bibfield  {journal} {\bibinfo  {journal} {Nature Photonics}\ }\textbf {\bibinfo {volume} {14}},\ \bibinfo {pages} {273} (\bibinfo {year} {2020})}\BibitemShut {NoStop}%
\bibitem [{\citenamefont {Piekarek}\ \emph {et~al.}(2017)\citenamefont {Piekarek}, \citenamefont {Bonneau}, \citenamefont {Miki}, \citenamefont {Yamashita}, \citenamefont {Fujiwara}, \citenamefont {Sasaki}, \citenamefont {Terai}, \citenamefont {Tanner}, \citenamefont {Natarajan}, \citenamefont {Hadfield} \emph {et~al.}}]{piekarek2017high}%
  \BibitemOpen
  \bibfield  {author} {\bibinfo {author} {\bibfnamefont {M.}~\bibnamefont {Piekarek}}, \bibinfo {author} {\bibfnamefont {D.}~\bibnamefont {Bonneau}}, \bibinfo {author} {\bibfnamefont {S.}~\bibnamefont {Miki}}, \bibinfo {author} {\bibfnamefont {T.}~\bibnamefont {Yamashita}}, \bibinfo {author} {\bibfnamefont {M.}~\bibnamefont {Fujiwara}}, \bibinfo {author} {\bibfnamefont {M.}~\bibnamefont {Sasaki}}, \bibinfo {author} {\bibfnamefont {H.}~\bibnamefont {Terai}}, \bibinfo {author} {\bibfnamefont {M.~G.}\ \bibnamefont {Tanner}}, \bibinfo {author} {\bibfnamefont {C.~M.}\ \bibnamefont {Natarajan}}, \bibinfo {author} {\bibfnamefont {R.~H.}\ \bibnamefont {Hadfield}}, \emph {et~al.},\ }\bibfield  {title} {\bibinfo {title} {High-extinction ratio integrated photonic filters for silicon quantum photonics},\ }\href@noop {} {\bibfield  {journal} {\bibinfo  {journal} {Optics letters}\ }\textbf {\bibinfo {volume} {42}},\ \bibinfo {pages} {815} (\bibinfo {year} {2017})}\BibitemShut {NoStop}%
\bibitem [{\citenamefont {Ong}\ \emph {et~al.}(2013)\citenamefont {Ong}, \citenamefont {Kumar},\ and\ \citenamefont {Mookherjea}}]{ong2013ultra}%
  \BibitemOpen
  \bibfield  {author} {\bibinfo {author} {\bibfnamefont {J.~R.}\ \bibnamefont {Ong}}, \bibinfo {author} {\bibfnamefont {R.}~\bibnamefont {Kumar}},\ and\ \bibinfo {author} {\bibfnamefont {S.}~\bibnamefont {Mookherjea}},\ }\bibfield  {title} {\bibinfo {title} {Ultra-high-contrast and tunable-bandwidth filter using cascaded high-order silicon microring filters},\ }\href@noop {} {\bibfield  {journal} {\bibinfo  {journal} {IEEE Photonics Technology Letters}\ }\textbf {\bibinfo {volume} {25}},\ \bibinfo {pages} {1543} (\bibinfo {year} {2013})}\BibitemShut {NoStop}%
\bibitem [{\citenamefont {Harris}\ \emph {et~al.}(2014)\citenamefont {Harris}, \citenamefont {Grassani}, \citenamefont {Simbula}, \citenamefont {Pant}, \citenamefont {Galli}, \citenamefont {Baehr-Jones}, \citenamefont {Hochberg}, \citenamefont {Englund}, \citenamefont {Bajoni},\ and\ \citenamefont {Galland}}]{harris2014integrated}%
  \BibitemOpen
  \bibfield  {author} {\bibinfo {author} {\bibfnamefont {N.~C.}\ \bibnamefont {Harris}}, \bibinfo {author} {\bibfnamefont {D.}~\bibnamefont {Grassani}}, \bibinfo {author} {\bibfnamefont {A.}~\bibnamefont {Simbula}}, \bibinfo {author} {\bibfnamefont {M.}~\bibnamefont {Pant}}, \bibinfo {author} {\bibfnamefont {M.}~\bibnamefont {Galli}}, \bibinfo {author} {\bibfnamefont {T.}~\bibnamefont {Baehr-Jones}}, \bibinfo {author} {\bibfnamefont {M.}~\bibnamefont {Hochberg}}, \bibinfo {author} {\bibfnamefont {D.}~\bibnamefont {Englund}}, \bibinfo {author} {\bibfnamefont {D.}~\bibnamefont {Bajoni}},\ and\ \bibinfo {author} {\bibfnamefont {C.}~\bibnamefont {Galland}},\ }\bibfield  {title} {\bibinfo {title} {Integrated source of spectrally filtered correlated photons for large-scale quantum photonic systems},\ }\href@noop {} {\bibfield  {journal} {\bibinfo  {journal} {Physical Review X}\ }\textbf {\bibinfo {volume} {4}},\ \bibinfo {pages} {041047} (\bibinfo {year} {2014})}\BibitemShut {NoStop}%
\bibitem [{\citenamefont {Wang}\ \emph {et~al.}(2023)\citenamefont {Wang}, \citenamefont {Faurby}, \citenamefont {Ruf}, \citenamefont {Sund}, \citenamefont {Nielsen}, \citenamefont {Volet}, \citenamefont {Heck}, \citenamefont {Bart}, \citenamefont {Wieck}, \citenamefont {Ludwig} \emph {et~al.}}]{wang2023deterministic}%
  \BibitemOpen
  \bibfield  {author} {\bibinfo {author} {\bibfnamefont {Y.}~\bibnamefont {Wang}}, \bibinfo {author} {\bibfnamefont {C.~F.}\ \bibnamefont {Faurby}}, \bibinfo {author} {\bibfnamefont {F.}~\bibnamefont {Ruf}}, \bibinfo {author} {\bibfnamefont {P.~I.}\ \bibnamefont {Sund}}, \bibinfo {author} {\bibfnamefont {K.~H.}\ \bibnamefont {Nielsen}}, \bibinfo {author} {\bibfnamefont {N.}~\bibnamefont {Volet}}, \bibinfo {author} {\bibfnamefont {M.~J.}\ \bibnamefont {Heck}}, \bibinfo {author} {\bibfnamefont {N.}~\bibnamefont {Bart}}, \bibinfo {author} {\bibfnamefont {A.~D.}\ \bibnamefont {Wieck}}, \bibinfo {author} {\bibfnamefont {A.}~\bibnamefont {Ludwig}}, \emph {et~al.},\ }\bibfield  {title} {\bibinfo {title} {Deterministic photon source interfaced with a programmable silicon-nitride integrated circuit},\ }\href@noop {} {\bibfield  {journal} {\bibinfo  {journal} {arXiv preprint arXiv:2302.06282}\ } (\bibinfo {year} {2023})}\BibitemShut {NoStop}%
\bibitem [{\citenamefont {Slussarenko}\ and\ \citenamefont {Pryde}(2019)}]{slussarenko2019photonic}%
  \BibitemOpen
  \bibfield  {author} {\bibinfo {author} {\bibfnamefont {S.}~\bibnamefont {Slussarenko}}\ and\ \bibinfo {author} {\bibfnamefont {G.~J.}\ \bibnamefont {Pryde}},\ }\bibfield  {title} {\bibinfo {title} {Photonic quantum information processing: A concise review},\ }\href@noop {} {\bibfield  {journal} {\bibinfo  {journal} {Applied Physics Reviews}\ }\textbf {\bibinfo {volume} {6}} (\bibinfo {year} {2019})}\BibitemShut {NoStop}%
\bibitem [{\citenamefont {Zhu}\ \emph {et~al.}(2021)\citenamefont {Zhu}, \citenamefont {Shao}, \citenamefont {Yu}, \citenamefont {Cheng}, \citenamefont {Desiatov}, \citenamefont {Xin}, \citenamefont {Hu}, \citenamefont {Holzgrafe}, \citenamefont {Ghosh}, \citenamefont {Shams-Ansari} \emph {et~al.}}]{zhu2021integrated}%
  \BibitemOpen
  \bibfield  {author} {\bibinfo {author} {\bibfnamefont {D.}~\bibnamefont {Zhu}}, \bibinfo {author} {\bibfnamefont {L.}~\bibnamefont {Shao}}, \bibinfo {author} {\bibfnamefont {M.}~\bibnamefont {Yu}}, \bibinfo {author} {\bibfnamefont {R.}~\bibnamefont {Cheng}}, \bibinfo {author} {\bibfnamefont {B.}~\bibnamefont {Desiatov}}, \bibinfo {author} {\bibfnamefont {C.}~\bibnamefont {Xin}}, \bibinfo {author} {\bibfnamefont {Y.}~\bibnamefont {Hu}}, \bibinfo {author} {\bibfnamefont {J.}~\bibnamefont {Holzgrafe}}, \bibinfo {author} {\bibfnamefont {S.}~\bibnamefont {Ghosh}}, \bibinfo {author} {\bibfnamefont {A.}~\bibnamefont {Shams-Ansari}}, \emph {et~al.},\ }\bibfield  {title} {\bibinfo {title} {Integrated photonics on thin-film lithium niobate},\ }\href@noop {} {\bibfield  {journal} {\bibinfo  {journal} {Advances in Optics and Photonics}\ }\textbf {\bibinfo {volume} {13}},\ \bibinfo {pages} {242} (\bibinfo {year} {2021})}\BibitemShut {NoStop}%
\bibitem [{\citenamefont {Xin}\ \emph {et~al.}(2022)\citenamefont {Xin}, \citenamefont {Mishra}, \citenamefont {Chen}, \citenamefont {Zhu}, \citenamefont {Shams-Ansari}, \citenamefont {Langrock}, \citenamefont {Sinclair}, \citenamefont {Wong}, \citenamefont {Fejer},\ and\ \citenamefont {Lon{\v{c}}ar}}]{xin2022spectrally}%
  \BibitemOpen
  \bibfield  {author} {\bibinfo {author} {\bibfnamefont {C.}~\bibnamefont {Xin}}, \bibinfo {author} {\bibfnamefont {J.}~\bibnamefont {Mishra}}, \bibinfo {author} {\bibfnamefont {C.}~\bibnamefont {Chen}}, \bibinfo {author} {\bibfnamefont {D.}~\bibnamefont {Zhu}}, \bibinfo {author} {\bibfnamefont {A.}~\bibnamefont {Shams-Ansari}}, \bibinfo {author} {\bibfnamefont {C.}~\bibnamefont {Langrock}}, \bibinfo {author} {\bibfnamefont {N.}~\bibnamefont {Sinclair}}, \bibinfo {author} {\bibfnamefont {F.~N.}\ \bibnamefont {Wong}}, \bibinfo {author} {\bibfnamefont {M.}~\bibnamefont {Fejer}},\ and\ \bibinfo {author} {\bibfnamefont {M.}~\bibnamefont {Lon{\v{c}}ar}},\ }\bibfield  {title} {\bibinfo {title} {Spectrally separable photon-pair generation in dispersion engineered thin-film lithium niobate},\ }\href@noop {} {\bibfield  {journal} {\bibinfo  {journal} {Optics Letters}\ }\textbf {\bibinfo {volume} {47}},\ \bibinfo {pages} {2830} (\bibinfo {year} {2022})}\BibitemShut {NoStop}%
\bibitem [{\citenamefont {Lomonte}\ \emph {et~al.}(2021{\natexlab{a}})\citenamefont {Lomonte}, \citenamefont {Wolff}, \citenamefont {Beutel}, \citenamefont {Ferrari}, \citenamefont {Schuck}, \citenamefont {Pernice},\ and\ \citenamefont {Lenzini}}]{lomonte2021single}%
  \BibitemOpen
  \bibfield  {author} {\bibinfo {author} {\bibfnamefont {E.}~\bibnamefont {Lomonte}}, \bibinfo {author} {\bibfnamefont {M.~A.}\ \bibnamefont {Wolff}}, \bibinfo {author} {\bibfnamefont {F.}~\bibnamefont {Beutel}}, \bibinfo {author} {\bibfnamefont {S.}~\bibnamefont {Ferrari}}, \bibinfo {author} {\bibfnamefont {C.}~\bibnamefont {Schuck}}, \bibinfo {author} {\bibfnamefont {W.~H.}\ \bibnamefont {Pernice}},\ and\ \bibinfo {author} {\bibfnamefont {F.}~\bibnamefont {Lenzini}},\ }\bibfield  {title} {\bibinfo {title} {Single-photon detection and cryogenic reconfigurability in lithium niobate nanophotonic circuits},\ }\href@noop {} {\bibfield  {journal} {\bibinfo  {journal} {Nature communications}\ }\textbf {\bibinfo {volume} {12}},\ \bibinfo {pages} {1} (\bibinfo {year} {2021}{\natexlab{a}})}\BibitemShut {NoStop}%
\bibitem [{\citenamefont {Eltes}\ \emph {et~al.}(2020)\citenamefont {Eltes}, \citenamefont {Villarreal-Garcia}, \citenamefont {Caimi}, \citenamefont {Siegwart}, \citenamefont {Gentile}, \citenamefont {Hart}, \citenamefont {Stark}, \citenamefont {Marshall}, \citenamefont {Thompson}, \citenamefont {Barreto} \emph {et~al.}}]{eltes2020integrated}%
  \BibitemOpen
  \bibfield  {author} {\bibinfo {author} {\bibfnamefont {F.}~\bibnamefont {Eltes}}, \bibinfo {author} {\bibfnamefont {G.~E.}\ \bibnamefont {Villarreal-Garcia}}, \bibinfo {author} {\bibfnamefont {D.}~\bibnamefont {Caimi}}, \bibinfo {author} {\bibfnamefont {H.}~\bibnamefont {Siegwart}}, \bibinfo {author} {\bibfnamefont {A.~A.}\ \bibnamefont {Gentile}}, \bibinfo {author} {\bibfnamefont {A.}~\bibnamefont {Hart}}, \bibinfo {author} {\bibfnamefont {P.}~\bibnamefont {Stark}}, \bibinfo {author} {\bibfnamefont {G.~D.}\ \bibnamefont {Marshall}}, \bibinfo {author} {\bibfnamefont {M.~G.}\ \bibnamefont {Thompson}}, \bibinfo {author} {\bibfnamefont {J.}~\bibnamefont {Barreto}}, \emph {et~al.},\ }\bibfield  {title} {\bibinfo {title} {An integrated optical modulator operating at cryogenic temperatures},\ }\href@noop {} {\bibfield  {journal} {\bibinfo  {journal} {Nature Materials}\ }\textbf {\bibinfo {volume} {19}},\ \bibinfo {pages} {1164} (\bibinfo {year} {2020})}\BibitemShut {NoStop}%
\bibitem [{\citenamefont {Wang}\ \emph {et~al.}(2018{\natexlab{b}})\citenamefont {Wang}, \citenamefont {Zhang}, \citenamefont {Chen}, \citenamefont {Bertrand}, \citenamefont {Shams-Ansari}, \citenamefont {Chandrasekhar}, \citenamefont {Winzer},\ and\ \citenamefont {Lon{\v{c}}ar}}]{wang2018integrated}%
  \BibitemOpen
  \bibfield  {author} {\bibinfo {author} {\bibfnamefont {C.}~\bibnamefont {Wang}}, \bibinfo {author} {\bibfnamefont {M.}~\bibnamefont {Zhang}}, \bibinfo {author} {\bibfnamefont {X.}~\bibnamefont {Chen}}, \bibinfo {author} {\bibfnamefont {M.}~\bibnamefont {Bertrand}}, \bibinfo {author} {\bibfnamefont {A.}~\bibnamefont {Shams-Ansari}}, \bibinfo {author} {\bibfnamefont {S.}~\bibnamefont {Chandrasekhar}}, \bibinfo {author} {\bibfnamefont {P.}~\bibnamefont {Winzer}},\ and\ \bibinfo {author} {\bibfnamefont {M.}~\bibnamefont {Lon{\v{c}}ar}},\ }\bibfield  {title} {\bibinfo {title} {Integrated lithium niobate electro-optic modulators operating at cmos-compatible voltages},\ }\href@noop {} {\bibfield  {journal} {\bibinfo  {journal} {Nature}\ }\textbf {\bibinfo {volume} {562}},\ \bibinfo {pages} {101} (\bibinfo {year} {2018}{\natexlab{b}})}\BibitemShut {NoStop}%
\bibitem [{\citenamefont {Xu}\ \emph {et~al.}(2020)\citenamefont {Xu}, \citenamefont {He}, \citenamefont {Zhang}, \citenamefont {Jian}, \citenamefont {Pan}, \citenamefont {Liu}, \citenamefont {Chen}, \citenamefont {Meng}, \citenamefont {Chen}, \citenamefont {Li} \emph {et~al.}}]{xu2020high}%
  \BibitemOpen
  \bibfield  {author} {\bibinfo {author} {\bibfnamefont {M.}~\bibnamefont {Xu}}, \bibinfo {author} {\bibfnamefont {M.}~\bibnamefont {He}}, \bibinfo {author} {\bibfnamefont {H.}~\bibnamefont {Zhang}}, \bibinfo {author} {\bibfnamefont {J.}~\bibnamefont {Jian}}, \bibinfo {author} {\bibfnamefont {Y.}~\bibnamefont {Pan}}, \bibinfo {author} {\bibfnamefont {X.}~\bibnamefont {Liu}}, \bibinfo {author} {\bibfnamefont {L.}~\bibnamefont {Chen}}, \bibinfo {author} {\bibfnamefont {X.}~\bibnamefont {Meng}}, \bibinfo {author} {\bibfnamefont {H.}~\bibnamefont {Chen}}, \bibinfo {author} {\bibfnamefont {Z.}~\bibnamefont {Li}}, \emph {et~al.},\ }\bibfield  {title} {\bibinfo {title} {High-performance coherent optical modulators based on thin-film lithium niobate platform},\ }\href@noop {} {\bibfield  {journal} {\bibinfo  {journal} {Nature communications}\ }\textbf {\bibinfo {volume} {11}},\ \bibinfo {pages} {3911} (\bibinfo {year} {2020})}\BibitemShut {NoStop}%
\bibitem [{\citenamefont {Zhang}\ \emph {et~al.}(2021)\citenamefont {Zhang}, \citenamefont {Wang}, \citenamefont {Kharel}, \citenamefont {Zhu},\ and\ \citenamefont {Lon{\v{c}}ar}}]{zhang2021integrated}%
  \BibitemOpen
  \bibfield  {author} {\bibinfo {author} {\bibfnamefont {M.}~\bibnamefont {Zhang}}, \bibinfo {author} {\bibfnamefont {C.}~\bibnamefont {Wang}}, \bibinfo {author} {\bibfnamefont {P.}~\bibnamefont {Kharel}}, \bibinfo {author} {\bibfnamefont {D.}~\bibnamefont {Zhu}},\ and\ \bibinfo {author} {\bibfnamefont {M.}~\bibnamefont {Lon{\v{c}}ar}},\ }\bibfield  {title} {\bibinfo {title} {Integrated lithium niobate electro-optic modulators: when performance meets scalability},\ }\href@noop {} {\bibfield  {journal} {\bibinfo  {journal} {Optica}\ }\textbf {\bibinfo {volume} {8}},\ \bibinfo {pages} {652} (\bibinfo {year} {2021})}\BibitemShut {NoStop}%
\bibitem [{\citenamefont {Wang}\ \emph {et~al.}(2019)\citenamefont {Wang}, \citenamefont {Zhang}, \citenamefont {Yu}, \citenamefont {Zhu}, \citenamefont {Hu},\ and\ \citenamefont {Loncar}}]{wang2019monolithic}%
  \BibitemOpen
  \bibfield  {author} {\bibinfo {author} {\bibfnamefont {C.}~\bibnamefont {Wang}}, \bibinfo {author} {\bibfnamefont {M.}~\bibnamefont {Zhang}}, \bibinfo {author} {\bibfnamefont {M.}~\bibnamefont {Yu}}, \bibinfo {author} {\bibfnamefont {R.}~\bibnamefont {Zhu}}, \bibinfo {author} {\bibfnamefont {H.}~\bibnamefont {Hu}},\ and\ \bibinfo {author} {\bibfnamefont {M.}~\bibnamefont {Loncar}},\ }\bibfield  {title} {\bibinfo {title} {Monolithic lithium niobate photonic circuits for kerr frequency comb generation and modulation},\ }\href@noop {} {\bibfield  {journal} {\bibinfo  {journal} {Nature communications}\ }\textbf {\bibinfo {volume} {10}},\ \bibinfo {pages} {978} (\bibinfo {year} {2019})}\BibitemShut {NoStop}%
\bibitem [{\citenamefont {Wang}\ \emph {et~al.}(2018{\natexlab{c}})\citenamefont {Wang}, \citenamefont {Langrock}, \citenamefont {Marandi}, \citenamefont {Jankowski}, \citenamefont {Zhang}, \citenamefont {Desiatov}, \citenamefont {Fejer},\ and\ \citenamefont {Lon{\v{c}}ar}}]{wang2018ultrahigh}%
  \BibitemOpen
  \bibfield  {author} {\bibinfo {author} {\bibfnamefont {C.}~\bibnamefont {Wang}}, \bibinfo {author} {\bibfnamefont {C.}~\bibnamefont {Langrock}}, \bibinfo {author} {\bibfnamefont {A.}~\bibnamefont {Marandi}}, \bibinfo {author} {\bibfnamefont {M.}~\bibnamefont {Jankowski}}, \bibinfo {author} {\bibfnamefont {M.}~\bibnamefont {Zhang}}, \bibinfo {author} {\bibfnamefont {B.}~\bibnamefont {Desiatov}}, \bibinfo {author} {\bibfnamefont {M.~M.}\ \bibnamefont {Fejer}},\ and\ \bibinfo {author} {\bibfnamefont {M.}~\bibnamefont {Lon{\v{c}}ar}},\ }\bibfield  {title} {\bibinfo {title} {Ultrahigh-efficiency wavelength conversion in nanophotonic periodically poled lithium niobate waveguides},\ }\href@noop {} {\bibfield  {journal} {\bibinfo  {journal} {Optica}\ }\textbf {\bibinfo {volume} {5}},\ \bibinfo {pages} {1438} (\bibinfo {year} {2018}{\natexlab{c}})}\BibitemShut {NoStop}%
\bibitem [{\citenamefont {Lu}\ \emph {et~al.}(2021)\citenamefont {Lu}, \citenamefont {Al~Sayem}, \citenamefont {Gong}, \citenamefont {Surya}, \citenamefont {Zou},\ and\ \citenamefont {Tang}}]{lu2021ultralow}%
  \BibitemOpen
  \bibfield  {author} {\bibinfo {author} {\bibfnamefont {J.}~\bibnamefont {Lu}}, \bibinfo {author} {\bibfnamefont {A.}~\bibnamefont {Al~Sayem}}, \bibinfo {author} {\bibfnamefont {Z.}~\bibnamefont {Gong}}, \bibinfo {author} {\bibfnamefont {J.~B.}\ \bibnamefont {Surya}}, \bibinfo {author} {\bibfnamefont {C.-L.}\ \bibnamefont {Zou}},\ and\ \bibinfo {author} {\bibfnamefont {H.~X.}\ \bibnamefont {Tang}},\ }\bibfield  {title} {\bibinfo {title} {Ultralow-threshold thin-film lithium niobate optical parametric oscillator},\ }\href@noop {} {\bibfield  {journal} {\bibinfo  {journal} {Optica}\ }\textbf {\bibinfo {volume} {8}},\ \bibinfo {pages} {539} (\bibinfo {year} {2021})}\BibitemShut {NoStop}%
\bibitem [{\citenamefont {Sayem}\ \emph {et~al.}(2021)\citenamefont {Sayem}, \citenamefont {Wang}, \citenamefont {Lu}, \citenamefont {Liu}, \citenamefont {Bruch},\ and\ \citenamefont {Tang}}]{sayem2021efficient}%
  \BibitemOpen
  \bibfield  {author} {\bibinfo {author} {\bibfnamefont {A.~A.}\ \bibnamefont {Sayem}}, \bibinfo {author} {\bibfnamefont {Y.}~\bibnamefont {Wang}}, \bibinfo {author} {\bibfnamefont {J.}~\bibnamefont {Lu}}, \bibinfo {author} {\bibfnamefont {X.}~\bibnamefont {Liu}}, \bibinfo {author} {\bibfnamefont {A.~W.}\ \bibnamefont {Bruch}},\ and\ \bibinfo {author} {\bibfnamefont {H.~X.}\ \bibnamefont {Tang}},\ }\bibfield  {title} {\bibinfo {title} {Efficient and tunable blue light generation using lithium niobate nonlinear photonics},\ }\href@noop {} {\bibfield  {journal} {\bibinfo  {journal} {Applied Physics Letters}\ }\textbf {\bibinfo {volume} {119}},\ \bibinfo {pages} {231104} (\bibinfo {year} {2021})}\BibitemShut {NoStop}%
\bibitem [{\citenamefont {Kashiwazaki}\ \emph {et~al.}(2020)\citenamefont {Kashiwazaki}, \citenamefont {Takanashi}, \citenamefont {Yamashima}, \citenamefont {Kazama}, \citenamefont {Enbutsu}, \citenamefont {Kasahara}, \citenamefont {Umeki},\ and\ \citenamefont {Furusawa}}]{kashiwazaki2020continuous}%
  \BibitemOpen
  \bibfield  {author} {\bibinfo {author} {\bibfnamefont {T.}~\bibnamefont {Kashiwazaki}}, \bibinfo {author} {\bibfnamefont {N.}~\bibnamefont {Takanashi}}, \bibinfo {author} {\bibfnamefont {T.}~\bibnamefont {Yamashima}}, \bibinfo {author} {\bibfnamefont {T.}~\bibnamefont {Kazama}}, \bibinfo {author} {\bibfnamefont {K.}~\bibnamefont {Enbutsu}}, \bibinfo {author} {\bibfnamefont {R.}~\bibnamefont {Kasahara}}, \bibinfo {author} {\bibfnamefont {T.}~\bibnamefont {Umeki}},\ and\ \bibinfo {author} {\bibfnamefont {A.}~\bibnamefont {Furusawa}},\ }\bibfield  {title} {\bibinfo {title} {Continuous-wave 6-db-squeezed light with 2.5-thz-bandwidth from single-mode ppln waveguide},\ }\href@noop {} {\bibfield  {journal} {\bibinfo  {journal} {APL Photonics}\ }\textbf {\bibinfo {volume} {5}} (\bibinfo {year} {2020})}\BibitemShut {NoStop}%
\bibitem [{\citenamefont {Ma}\ \emph {et~al.}(2020)\citenamefont {Ma}, \citenamefont {Chen}, \citenamefont {Li}, \citenamefont {Tang}, \citenamefont {Sua}, \citenamefont {Fan},\ and\ \citenamefont {Huang}}]{ma2020ultrabright}%
  \BibitemOpen
  \bibfield  {author} {\bibinfo {author} {\bibfnamefont {Z.}~\bibnamefont {Ma}}, \bibinfo {author} {\bibfnamefont {J.-Y.}\ \bibnamefont {Chen}}, \bibinfo {author} {\bibfnamefont {Z.}~\bibnamefont {Li}}, \bibinfo {author} {\bibfnamefont {C.}~\bibnamefont {Tang}}, \bibinfo {author} {\bibfnamefont {Y.~M.}\ \bibnamefont {Sua}}, \bibinfo {author} {\bibfnamefont {H.}~\bibnamefont {Fan}},\ and\ \bibinfo {author} {\bibfnamefont {Y.-P.}\ \bibnamefont {Huang}},\ }\bibfield  {title} {\bibinfo {title} {Ultrabright quantum photon sources on chip},\ }\href@noop {} {\bibfield  {journal} {\bibinfo  {journal} {Physical Review Letters}\ }\textbf {\bibinfo {volume} {125}},\ \bibinfo {pages} {263602} (\bibinfo {year} {2020})}\BibitemShut {NoStop}%
\bibitem [{\citenamefont {Chen}\ \emph {et~al.}(2022)\citenamefont {Chen}, \citenamefont {Briggs}, \citenamefont {Hou},\ and\ \citenamefont {Fan}}]{chen2022ultra}%
  \BibitemOpen
  \bibfield  {author} {\bibinfo {author} {\bibfnamefont {P.-K.}\ \bibnamefont {Chen}}, \bibinfo {author} {\bibfnamefont {I.}~\bibnamefont {Briggs}}, \bibinfo {author} {\bibfnamefont {S.}~\bibnamefont {Hou}},\ and\ \bibinfo {author} {\bibfnamefont {L.}~\bibnamefont {Fan}},\ }\bibfield  {title} {\bibinfo {title} {Ultra-broadband quadrature squeezing with thin-film lithium niobate nanophotonics},\ }\href@noop {} {\bibfield  {journal} {\bibinfo  {journal} {Optics Letters}\ }\textbf {\bibinfo {volume} {47}},\ \bibinfo {pages} {1506} (\bibinfo {year} {2022})}\BibitemShut {NoStop}%
\bibitem [{\citenamefont {Nehra}\ \emph {et~al.}(2022)\citenamefont {Nehra}, \citenamefont {Sekine}, \citenamefont {Ledezma}, \citenamefont {Guo}, \citenamefont {Gray}, \citenamefont {Roy},\ and\ \citenamefont {Marandi}}]{nehra2022few}%
  \BibitemOpen
  \bibfield  {author} {\bibinfo {author} {\bibfnamefont {R.}~\bibnamefont {Nehra}}, \bibinfo {author} {\bibfnamefont {R.}~\bibnamefont {Sekine}}, \bibinfo {author} {\bibfnamefont {L.}~\bibnamefont {Ledezma}}, \bibinfo {author} {\bibfnamefont {Q.}~\bibnamefont {Guo}}, \bibinfo {author} {\bibfnamefont {R.~M.}\ \bibnamefont {Gray}}, \bibinfo {author} {\bibfnamefont {A.}~\bibnamefont {Roy}},\ and\ \bibinfo {author} {\bibfnamefont {A.}~\bibnamefont {Marandi}},\ }\bibfield  {title} {\bibinfo {title} {Few-cycle vacuum squeezing in nanophotonics},\ }\href@noop {} {\bibfield  {journal} {\bibinfo  {journal} {Science}\ }\textbf {\bibinfo {volume} {377}},\ \bibinfo {pages} {1333} (\bibinfo {year} {2022})}\BibitemShut {NoStop}%
\bibitem [{\citenamefont {Sayem}\ \emph {et~al.}(2020)\citenamefont {Sayem}, \citenamefont {Cheng}, \citenamefont {Wang},\ and\ \citenamefont {Tang}}]{sayem2020lithium}%
  \BibitemOpen
  \bibfield  {author} {\bibinfo {author} {\bibfnamefont {A.~A.}\ \bibnamefont {Sayem}}, \bibinfo {author} {\bibfnamefont {R.}~\bibnamefont {Cheng}}, \bibinfo {author} {\bibfnamefont {S.}~\bibnamefont {Wang}},\ and\ \bibinfo {author} {\bibfnamefont {H.~X.}\ \bibnamefont {Tang}},\ }\bibfield  {title} {\bibinfo {title} {Lithium-niobate-on-insulator waveguide-integrated superconducting nanowire single-photon detectors},\ }\href@noop {} {\bibfield  {journal} {\bibinfo  {journal} {Applied Physics Letters}\ }\textbf {\bibinfo {volume} {116}},\ \bibinfo {pages} {151102} (\bibinfo {year} {2020})}\BibitemShut {NoStop}%
\bibitem [{\citenamefont {Guo}\ \emph {et~al.}(2016)\citenamefont {Guo}, \citenamefont {Zou},\ and\ \citenamefont {Tang}}]{guo201670}%
  \BibitemOpen
  \bibfield  {author} {\bibinfo {author} {\bibfnamefont {X.}~\bibnamefont {Guo}}, \bibinfo {author} {\bibfnamefont {C.-L.}\ \bibnamefont {Zou}},\ and\ \bibinfo {author} {\bibfnamefont {H.~X.}\ \bibnamefont {Tang}},\ }\bibfield  {title} {\bibinfo {title} {70 db long-pass filter on a nanophotonic chip},\ }\href@noop {} {\bibfield  {journal} {\bibinfo  {journal} {Optics express}\ }\textbf {\bibinfo {volume} {24}},\ \bibinfo {pages} {21167} (\bibinfo {year} {2016})}\BibitemShut {NoStop}%
\bibitem [{\citenamefont {Sayem}(2023)}]{Sayemthesis}%
  \BibitemOpen
  \bibfield  {author} {\bibinfo {author} {\bibfnamefont {A.~A.}\ \bibnamefont {Sayem}},\ }\emph {\bibinfo {title} {Integrated Nano-Photonic Circuits on Thin-Film Lithium Niobate}},\ \href {https://www.proquest.com/dissertations-theses/integrated-nano-photonic-circuits-on-thin-film/docview/2834863316/se-2} {Ph.D. thesis} (\bibinfo {year} {2023})\BibitemShut {NoStop}%
\bibitem [{\citenamefont {Takagi}\ \emph {et~al.}(1996)\citenamefont {Takagi}, \citenamefont {Kikuchi}, \citenamefont {Maeda}, \citenamefont {Chung},\ and\ \citenamefont {Suga}}]{takagi1996surface}%
  \BibitemOpen
  \bibfield  {author} {\bibinfo {author} {\bibfnamefont {H.}~\bibnamefont {Takagi}}, \bibinfo {author} {\bibfnamefont {K.}~\bibnamefont {Kikuchi}}, \bibinfo {author} {\bibfnamefont {R.}~\bibnamefont {Maeda}}, \bibinfo {author} {\bibfnamefont {T.}~\bibnamefont {Chung}},\ and\ \bibinfo {author} {\bibfnamefont {T.}~\bibnamefont {Suga}},\ }\bibfield  {title} {\bibinfo {title} {Surface activated bonding of silicon wafers at room temperature},\ }\href@noop {} {\bibfield  {journal} {\bibinfo  {journal} {Applied physics letters}\ }\textbf {\bibinfo {volume} {68}},\ \bibinfo {pages} {2222} (\bibinfo {year} {1996})}\BibitemShut {NoStop}%
\bibitem [{\citenamefont {Takakura}\ \emph {et~al.}(2023)\citenamefont {Takakura}, \citenamefont {Murakami}, \citenamefont {Watanabe},\ and\ \citenamefont {Takigawa}}]{takakura2023room}%
  \BibitemOpen
  \bibfield  {author} {\bibinfo {author} {\bibfnamefont {R.}~\bibnamefont {Takakura}}, \bibinfo {author} {\bibfnamefont {S.}~\bibnamefont {Murakami}}, \bibinfo {author} {\bibfnamefont {K.}~\bibnamefont {Watanabe}},\ and\ \bibinfo {author} {\bibfnamefont {R.}~\bibnamefont {Takigawa}},\ }\bibfield  {title} {\bibinfo {title} {Room-temperature bonding of al2o3 thin films deposited using atomic layer deposition},\ }\href@noop {} {\bibfield  {journal} {\bibinfo  {journal} {Scientific reports}\ }\textbf {\bibinfo {volume} {13}},\ \bibinfo {pages} {3581} (\bibinfo {year} {2023})}\BibitemShut {NoStop}%
\bibitem [{\citenamefont {Churaev}\ \emph {et~al.}(2023)\citenamefont {Churaev}, \citenamefont {Wang}, \citenamefont {Riedhauser}, \citenamefont {Snigirev}, \citenamefont {Bl{\'e}sin}, \citenamefont {M{\"o}hl}, \citenamefont {Anderson}, \citenamefont {Siddharth}, \citenamefont {Popoff}, \citenamefont {Drechsler} \emph {et~al.}}]{churaev2023heterogeneously}%
  \BibitemOpen
  \bibfield  {author} {\bibinfo {author} {\bibfnamefont {M.}~\bibnamefont {Churaev}}, \bibinfo {author} {\bibfnamefont {R.~N.}\ \bibnamefont {Wang}}, \bibinfo {author} {\bibfnamefont {A.}~\bibnamefont {Riedhauser}}, \bibinfo {author} {\bibfnamefont {V.}~\bibnamefont {Snigirev}}, \bibinfo {author} {\bibfnamefont {T.}~\bibnamefont {Bl{\'e}sin}}, \bibinfo {author} {\bibfnamefont {C.}~\bibnamefont {M{\"o}hl}}, \bibinfo {author} {\bibfnamefont {M.~H.}\ \bibnamefont {Anderson}}, \bibinfo {author} {\bibfnamefont {A.}~\bibnamefont {Siddharth}}, \bibinfo {author} {\bibfnamefont {Y.}~\bibnamefont {Popoff}}, \bibinfo {author} {\bibfnamefont {U.}~\bibnamefont {Drechsler}}, \emph {et~al.},\ }\bibfield  {title} {\bibinfo {title} {A heterogeneously integrated lithium niobate-on-silicon nitride photonic platform},\ }\href@noop {} {\bibfield  {journal} {\bibinfo  {journal} {Nature Communications}\ }\textbf {\bibinfo {volume} {14}},\ \bibinfo {pages} {3499} (\bibinfo {year} {2023})}\BibitemShut {NoStop}%
\bibitem [{\citenamefont {Lin}\ \emph {et~al.}(2022)\citenamefont {Lin}, \citenamefont {Lin}, \citenamefont {Li}, \citenamefont {Xu}, \citenamefont {He}, \citenamefont {Ke}, \citenamefont {Tan}, \citenamefont {Han}, \citenamefont {Li}, \citenamefont {Wang} \emph {et~al.}}]{lin2022high}%
  \BibitemOpen
  \bibfield  {author} {\bibinfo {author} {\bibfnamefont {Z.}~\bibnamefont {Lin}}, \bibinfo {author} {\bibfnamefont {Y.}~\bibnamefont {Lin}}, \bibinfo {author} {\bibfnamefont {H.}~\bibnamefont {Li}}, \bibinfo {author} {\bibfnamefont {M.}~\bibnamefont {Xu}}, \bibinfo {author} {\bibfnamefont {M.}~\bibnamefont {He}}, \bibinfo {author} {\bibfnamefont {W.}~\bibnamefont {Ke}}, \bibinfo {author} {\bibfnamefont {H.}~\bibnamefont {Tan}}, \bibinfo {author} {\bibfnamefont {Y.}~\bibnamefont {Han}}, \bibinfo {author} {\bibfnamefont {Z.}~\bibnamefont {Li}}, \bibinfo {author} {\bibfnamefont {D.}~\bibnamefont {Wang}}, \emph {et~al.},\ }\bibfield  {title} {\bibinfo {title} {High-performance polarization management devices based on thin-film lithium niobate},\ }\href@noop {} {\bibfield  {journal} {\bibinfo  {journal} {Light: Science \& Applications}\ }\textbf {\bibinfo {volume} {11}},\ \bibinfo {pages} {93} (\bibinfo {year} {2022})}\BibitemShut {NoStop}%
\bibitem [{\citenamefont {Luo}\ \emph {et~al.}(2021)\citenamefont {Luo}, \citenamefont {Chen}, \citenamefont {Li}, \citenamefont {Chen}, \citenamefont {Han}, \citenamefont {Lin}, \citenamefont {Yu},\ and\ \citenamefont {Cai}}]{luo2021high}%
  \BibitemOpen
  \bibfield  {author} {\bibinfo {author} {\bibfnamefont {H.}~\bibnamefont {Luo}}, \bibinfo {author} {\bibfnamefont {Z.}~\bibnamefont {Chen}}, \bibinfo {author} {\bibfnamefont {H.}~\bibnamefont {Li}}, \bibinfo {author} {\bibfnamefont {L.}~\bibnamefont {Chen}}, \bibinfo {author} {\bibfnamefont {Y.}~\bibnamefont {Han}}, \bibinfo {author} {\bibfnamefont {Z.}~\bibnamefont {Lin}}, \bibinfo {author} {\bibfnamefont {S.}~\bibnamefont {Yu}},\ and\ \bibinfo {author} {\bibfnamefont {X.}~\bibnamefont {Cai}},\ }\bibfield  {title} {\bibinfo {title} {High-performance polarization splitter-rotator based on lithium niobate-on-insulator platform},\ }\href@noop {} {\bibfield  {journal} {\bibinfo  {journal} {IEEE Photonics Technology Letters}\ }\textbf {\bibinfo {volume} {33}},\ \bibinfo {pages} {1423} (\bibinfo {year} {2021})}\BibitemShut {NoStop}%
\bibitem [{\citenamefont {Bludau}\ \emph {et~al.}(1974)\citenamefont {Bludau}, \citenamefont {Onton},\ and\ \citenamefont {Heinke}}]{bludau1974temperature}%
  \BibitemOpen
  \bibfield  {author} {\bibinfo {author} {\bibfnamefont {W.}~\bibnamefont {Bludau}}, \bibinfo {author} {\bibfnamefont {A.}~\bibnamefont {Onton}},\ and\ \bibinfo {author} {\bibfnamefont {W.}~\bibnamefont {Heinke}},\ }\bibfield  {title} {\bibinfo {title} {Temperature dependence of the band gap of silicon},\ }\href@noop {} {\bibfield  {journal} {\bibinfo  {journal} {Journal of Applied Physics}\ }\textbf {\bibinfo {volume} {45}},\ \bibinfo {pages} {1846} (\bibinfo {year} {1974})}\BibitemShut {NoStop}%
\bibitem [{\citenamefont {Nguyen}\ \emph {et~al.}(2014)\citenamefont {Nguyen}, \citenamefont {Rougieux}, \citenamefont {Mitchell},\ and\ \citenamefont {Macdonald}}]{nguyen2014temperature}%
  \BibitemOpen
  \bibfield  {author} {\bibinfo {author} {\bibfnamefont {H.~T.}\ \bibnamefont {Nguyen}}, \bibinfo {author} {\bibfnamefont {F.~E.}\ \bibnamefont {Rougieux}}, \bibinfo {author} {\bibfnamefont {B.}~\bibnamefont {Mitchell}},\ and\ \bibinfo {author} {\bibfnamefont {D.}~\bibnamefont {Macdonald}},\ }\bibfield  {title} {\bibinfo {title} {Temperature dependence of the band-band absorption coefficient in crystalline silicon from photoluminescence},\ }\href@noop {} {\bibfield  {journal} {\bibinfo  {journal} {Journal of Applied Physics}\ }\textbf {\bibinfo {volume} {115}},\ \bibinfo {pages} {043710} (\bibinfo {year} {2014})}\BibitemShut {NoStop}%
\bibitem [{\citenamefont {Noffsinger}\ \emph {et~al.}(2012)\citenamefont {Noffsinger}, \citenamefont {Kioupakis}, \citenamefont {Van~de Walle}, \citenamefont {Louie},\ and\ \citenamefont {Cohen}}]{noffsinger2012phonon}%
  \BibitemOpen
  \bibfield  {author} {\bibinfo {author} {\bibfnamefont {J.}~\bibnamefont {Noffsinger}}, \bibinfo {author} {\bibfnamefont {E.}~\bibnamefont {Kioupakis}}, \bibinfo {author} {\bibfnamefont {C.~G.}\ \bibnamefont {Van~de Walle}}, \bibinfo {author} {\bibfnamefont {S.~G.}\ \bibnamefont {Louie}},\ and\ \bibinfo {author} {\bibfnamefont {M.~L.}\ \bibnamefont {Cohen}},\ }\bibfield  {title} {\bibinfo {title} {Phonon-assisted optical absorption in silicon from first principles},\ }\href@noop {} {\bibfield  {journal} {\bibinfo  {journal} {Physical review letters}\ }\textbf {\bibinfo {volume} {108}},\ \bibinfo {pages} {167402} (\bibinfo {year} {2012})}\BibitemShut {NoStop}%
\bibitem [{\citenamefont {Lomonte}\ \emph {et~al.}(2021{\natexlab{b}})\citenamefont {Lomonte}, \citenamefont {Lenzini},\ and\ \citenamefont {Pernice}}]{lomonte2021efficient}%
  \BibitemOpen
  \bibfield  {author} {\bibinfo {author} {\bibfnamefont {E.}~\bibnamefont {Lomonte}}, \bibinfo {author} {\bibfnamefont {F.}~\bibnamefont {Lenzini}},\ and\ \bibinfo {author} {\bibfnamefont {W.~H.}\ \bibnamefont {Pernice}},\ }\bibfield  {title} {\bibinfo {title} {Efficient self-imaging grating couplers on a lithium-niobate-on-insulator platform at near-visible and telecom wavelengths},\ }\href@noop {} {\bibfield  {journal} {\bibinfo  {journal} {Optics Express}\ }\textbf {\bibinfo {volume} {29}},\ \bibinfo {pages} {20205} (\bibinfo {year} {2021}{\natexlab{b}})}\BibitemShut {NoStop}%
\bibitem [{\citenamefont {Anant}\ \emph {et~al.}(2008)\citenamefont {Anant}, \citenamefont {Kerman}, \citenamefont {Dauler}, \citenamefont {Yang}, \citenamefont {Rosfjord},\ and\ \citenamefont {Berggren}}]{anant2008optical}%
  \BibitemOpen
  \bibfield  {author} {\bibinfo {author} {\bibfnamefont {V.}~\bibnamefont {Anant}}, \bibinfo {author} {\bibfnamefont {A.~J.}\ \bibnamefont {Kerman}}, \bibinfo {author} {\bibfnamefont {E.~A.}\ \bibnamefont {Dauler}}, \bibinfo {author} {\bibfnamefont {J.~K.}\ \bibnamefont {Yang}}, \bibinfo {author} {\bibfnamefont {K.~M.}\ \bibnamefont {Rosfjord}},\ and\ \bibinfo {author} {\bibfnamefont {K.~K.}\ \bibnamefont {Berggren}},\ }\bibfield  {title} {\bibinfo {title} {Optical properties of superconducting nanowire single-photon detectors},\ }\href@noop {} {\bibfield  {journal} {\bibinfo  {journal} {Optics express}\ }\textbf {\bibinfo {volume} {16}},\ \bibinfo {pages} {10750} (\bibinfo {year} {2008})}\BibitemShut {NoStop}%
\bibitem [{\citenamefont {Rosfjord}\ \emph {et~al.}(2006)\citenamefont {Rosfjord}, \citenamefont {Yang}, \citenamefont {Dauler}, \citenamefont {Kerman}, \citenamefont {Anant}, \citenamefont {Voronov}, \citenamefont {Gol’Tsman},\ and\ \citenamefont {Berggren}}]{rosfjord2006nanowire}%
  \BibitemOpen
  \bibfield  {author} {\bibinfo {author} {\bibfnamefont {K.~M.}\ \bibnamefont {Rosfjord}}, \bibinfo {author} {\bibfnamefont {J.~K.}\ \bibnamefont {Yang}}, \bibinfo {author} {\bibfnamefont {E.~A.}\ \bibnamefont {Dauler}}, \bibinfo {author} {\bibfnamefont {A.~J.}\ \bibnamefont {Kerman}}, \bibinfo {author} {\bibfnamefont {V.}~\bibnamefont {Anant}}, \bibinfo {author} {\bibfnamefont {B.~M.}\ \bibnamefont {Voronov}}, \bibinfo {author} {\bibfnamefont {G.~N.}\ \bibnamefont {Gol’Tsman}},\ and\ \bibinfo {author} {\bibfnamefont {K.~K.}\ \bibnamefont {Berggren}},\ }\bibfield  {title} {\bibinfo {title} {Nanowire single-photon detector with an integrated optical cavity and anti-reflection coating},\ }\href@noop {} {\bibfield  {journal} {\bibinfo  {journal} {Optics express}\ }\textbf {\bibinfo {volume} {14}},\ \bibinfo {pages} {527} (\bibinfo {year} {2006})}\BibitemShut {NoStop}%
\bibitem [{\citenamefont {Tanner}\ \emph {et~al.}(2010)\citenamefont {Tanner}, \citenamefont {Natarajan}, \citenamefont {Pottapenjara}, \citenamefont {O’Connor}, \citenamefont {Warburton}, \citenamefont {Hadfield}, \citenamefont {Baek}, \citenamefont {Nam}, \citenamefont {Dorenbos}, \citenamefont {Ure{\~n}a} \emph {et~al.}}]{tanner2010enhanced}%
  \BibitemOpen
  \bibfield  {author} {\bibinfo {author} {\bibfnamefont {M.~G.}\ \bibnamefont {Tanner}}, \bibinfo {author} {\bibfnamefont {C.}~\bibnamefont {Natarajan}}, \bibinfo {author} {\bibfnamefont {V.}~\bibnamefont {Pottapenjara}}, \bibinfo {author} {\bibfnamefont {J.}~\bibnamefont {O’Connor}}, \bibinfo {author} {\bibfnamefont {R.}~\bibnamefont {Warburton}}, \bibinfo {author} {\bibfnamefont {R.}~\bibnamefont {Hadfield}}, \bibinfo {author} {\bibfnamefont {B.}~\bibnamefont {Baek}}, \bibinfo {author} {\bibfnamefont {S.}~\bibnamefont {Nam}}, \bibinfo {author} {\bibfnamefont {S.}~\bibnamefont {Dorenbos}}, \bibinfo {author} {\bibfnamefont {E.~B.}\ \bibnamefont {Ure{\~n}a}}, \emph {et~al.},\ }\bibfield  {title} {\bibinfo {title} {Enhanced telecom wavelength single-photon detection with nbtin superconducting nanowires on oxidized silicon},\ }\href@noop {} {\bibfield  {journal} {\bibinfo  {journal} {Applied Physics Letters}\ }\textbf {\bibinfo {volume} {96}} (\bibinfo {year} {2010})}\BibitemShut {NoStop}%
\bibitem [{\citenamefont {Li}\ \emph {et~al.}(2014)\citenamefont {Li}, \citenamefont {Zhang}, \citenamefont {You}, \citenamefont {Zhang}, \citenamefont {Yang}, \citenamefont {Liu}, \citenamefont {Chen}, \citenamefont {Lv}, \citenamefont {Peng}, \citenamefont {Wang} \emph {et~al.}}]{li2014nonideal}%
  \BibitemOpen
  \bibfield  {author} {\bibinfo {author} {\bibfnamefont {H.}~\bibnamefont {Li}}, \bibinfo {author} {\bibfnamefont {W.}~\bibnamefont {Zhang}}, \bibinfo {author} {\bibfnamefont {L.}~\bibnamefont {You}}, \bibinfo {author} {\bibfnamefont {L.}~\bibnamefont {Zhang}}, \bibinfo {author} {\bibfnamefont {X.}~\bibnamefont {Yang}}, \bibinfo {author} {\bibfnamefont {X.}~\bibnamefont {Liu}}, \bibinfo {author} {\bibfnamefont {S.}~\bibnamefont {Chen}}, \bibinfo {author} {\bibfnamefont {C.}~\bibnamefont {Lv}}, \bibinfo {author} {\bibfnamefont {W.}~\bibnamefont {Peng}}, \bibinfo {author} {\bibfnamefont {Z.}~\bibnamefont {Wang}}, \emph {et~al.},\ }\bibfield  {title} {\bibinfo {title} {Nonideal optical cavity structure of superconducting nanowire single-photon detector},\ }\href@noop {} {\bibfield  {journal} {\bibinfo  {journal} {IEEE Journal of Selected Topics in Quantum Electronics}\ }\textbf {\bibinfo {volume} {20}},\ \bibinfo {pages} {198} (\bibinfo {year} {2014})}\BibitemShut {NoStop}%
\bibitem [{\citenamefont {Reddy}\ \emph {et~al.}(2020)\citenamefont {Reddy}, \citenamefont {Nerem}, \citenamefont {Nam}, \citenamefont {Mirin},\ and\ \citenamefont {Verma}}]{reddy2020superconducting}%
  \BibitemOpen
  \bibfield  {author} {\bibinfo {author} {\bibfnamefont {D.~V.}\ \bibnamefont {Reddy}}, \bibinfo {author} {\bibfnamefont {R.~R.}\ \bibnamefont {Nerem}}, \bibinfo {author} {\bibfnamefont {S.~W.}\ \bibnamefont {Nam}}, \bibinfo {author} {\bibfnamefont {R.~P.}\ \bibnamefont {Mirin}},\ and\ \bibinfo {author} {\bibfnamefont {V.~B.}\ \bibnamefont {Verma}},\ }\bibfield  {title} {\bibinfo {title} {Superconducting nanowire single-photon detectors with 98\% system detection efficiency at 1550 nm},\ }\href@noop {} {\bibfield  {journal} {\bibinfo  {journal} {Optica}\ }\textbf {\bibinfo {volume} {7}},\ \bibinfo {pages} {1649} (\bibinfo {year} {2020})}\BibitemShut {NoStop}%
\bibitem [{\citenamefont {Cheng}\ \emph {et~al.}(2019)\citenamefont {Cheng}, \citenamefont {Wang},\ and\ \citenamefont {Tang}}]{cheng2019superconducting}%
  \BibitemOpen
  \bibfield  {author} {\bibinfo {author} {\bibfnamefont {R.}~\bibnamefont {Cheng}}, \bibinfo {author} {\bibfnamefont {S.}~\bibnamefont {Wang}},\ and\ \bibinfo {author} {\bibfnamefont {H.~X.}\ \bibnamefont {Tang}},\ }\bibfield  {title} {\bibinfo {title} {Superconducting nanowire single-photon detectors fabricated from atomic-layer-deposited nbn},\ }\href@noop {} {\bibfield  {journal} {\bibinfo  {journal} {Applied Physics Letters}\ }\textbf {\bibinfo {volume} {115}} (\bibinfo {year} {2019})}\BibitemShut {NoStop}%
\bibitem [{\citenamefont {Esmaeil~Zadeh}\ \emph {et~al.}(2021)\citenamefont {Esmaeil~Zadeh}, \citenamefont {Chang}, \citenamefont {Los}, \citenamefont {Gyger}, \citenamefont {Elshaari}, \citenamefont {Steinhauer}, \citenamefont {Dorenbos},\ and\ \citenamefont {Zwiller}}]{esmaeil2021superconducting}%
  \BibitemOpen
  \bibfield  {author} {\bibinfo {author} {\bibfnamefont {I.}~\bibnamefont {Esmaeil~Zadeh}}, \bibinfo {author} {\bibfnamefont {J.}~\bibnamefont {Chang}}, \bibinfo {author} {\bibfnamefont {J.~W.}\ \bibnamefont {Los}}, \bibinfo {author} {\bibfnamefont {S.}~\bibnamefont {Gyger}}, \bibinfo {author} {\bibfnamefont {A.~W.}\ \bibnamefont {Elshaari}}, \bibinfo {author} {\bibfnamefont {S.}~\bibnamefont {Steinhauer}}, \bibinfo {author} {\bibfnamefont {S.~N.}\ \bibnamefont {Dorenbos}},\ and\ \bibinfo {author} {\bibfnamefont {V.}~\bibnamefont {Zwiller}},\ }\bibfield  {title} {\bibinfo {title} {Superconducting nanowire single-photon detectors: A perspective on evolution, state-of-the-art, future developments, and applications},\ }\href@noop {} {\bibfield  {journal} {\bibinfo  {journal} {Applied Physics Letters}\ }\textbf {\bibinfo {volume} {118}} (\bibinfo {year} {2021})}\BibitemShut {NoStop}%
\bibitem [{\citenamefont {Natarajan}\ \emph {et~al.}(2012)\citenamefont {Natarajan}, \citenamefont {Tanner},\ and\ \citenamefont {Hadfield}}]{natarajan2012superconducting}%
  \BibitemOpen
  \bibfield  {author} {\bibinfo {author} {\bibfnamefont {C.~M.}\ \bibnamefont {Natarajan}}, \bibinfo {author} {\bibfnamefont {M.~G.}\ \bibnamefont {Tanner}},\ and\ \bibinfo {author} {\bibfnamefont {R.~H.}\ \bibnamefont {Hadfield}},\ }\bibfield  {title} {\bibinfo {title} {Superconducting nanowire single-photon detectors: physics and applications},\ }\href@noop {} {\bibfield  {journal} {\bibinfo  {journal} {Superconductor science and technology}\ }\textbf {\bibinfo {volume} {25}},\ \bibinfo {pages} {063001} (\bibinfo {year} {2012})}\BibitemShut {NoStop}%
\bibitem [{\citenamefont {Marsili}\ \emph {et~al.}(2013)\citenamefont {Marsili}, \citenamefont {Verma}, \citenamefont {Stern}, \citenamefont {Harrington}, \citenamefont {Lita}, \citenamefont {Gerrits}, \citenamefont {Vayshenker}, \citenamefont {Baek}, \citenamefont {Shaw}, \citenamefont {Mirin} \emph {et~al.}}]{marsili2013detecting}%
  \BibitemOpen
  \bibfield  {author} {\bibinfo {author} {\bibfnamefont {F.}~\bibnamefont {Marsili}}, \bibinfo {author} {\bibfnamefont {V.~B.}\ \bibnamefont {Verma}}, \bibinfo {author} {\bibfnamefont {J.~A.}\ \bibnamefont {Stern}}, \bibinfo {author} {\bibfnamefont {S.}~\bibnamefont {Harrington}}, \bibinfo {author} {\bibfnamefont {A.~E.}\ \bibnamefont {Lita}}, \bibinfo {author} {\bibfnamefont {T.}~\bibnamefont {Gerrits}}, \bibinfo {author} {\bibfnamefont {I.}~\bibnamefont {Vayshenker}}, \bibinfo {author} {\bibfnamefont {B.}~\bibnamefont {Baek}}, \bibinfo {author} {\bibfnamefont {M.~D.}\ \bibnamefont {Shaw}}, \bibinfo {author} {\bibfnamefont {R.~P.}\ \bibnamefont {Mirin}}, \emph {et~al.},\ }\bibfield  {title} {\bibinfo {title} {Detecting single infrared photons with 93\% system efficiency},\ }\href@noop {} {\bibfield  {journal} {\bibinfo  {journal} {Nature Photonics}\ }\textbf {\bibinfo {volume} {7}},\ \bibinfo {pages} {210} (\bibinfo {year} {2013})}\BibitemShut {NoStop}%
\bibitem [{\citenamefont {Verma}\ \emph {et~al.}(2015)\citenamefont {Verma}, \citenamefont {Korzh}, \citenamefont {Bussieres}, \citenamefont {Horansky}, \citenamefont {Dyer}, \citenamefont {Lita}, \citenamefont {Vayshenker}, \citenamefont {Marsili}, \citenamefont {Shaw}, \citenamefont {Zbinden} \emph {et~al.}}]{verma2015high}%
  \BibitemOpen
  \bibfield  {author} {\bibinfo {author} {\bibfnamefont {V.~B.}\ \bibnamefont {Verma}}, \bibinfo {author} {\bibfnamefont {B.}~\bibnamefont {Korzh}}, \bibinfo {author} {\bibfnamefont {F.}~\bibnamefont {Bussieres}}, \bibinfo {author} {\bibfnamefont {R.~D.}\ \bibnamefont {Horansky}}, \bibinfo {author} {\bibfnamefont {S.~D.}\ \bibnamefont {Dyer}}, \bibinfo {author} {\bibfnamefont {A.~E.}\ \bibnamefont {Lita}}, \bibinfo {author} {\bibfnamefont {I.}~\bibnamefont {Vayshenker}}, \bibinfo {author} {\bibfnamefont {F.}~\bibnamefont {Marsili}}, \bibinfo {author} {\bibfnamefont {M.~D.}\ \bibnamefont {Shaw}}, \bibinfo {author} {\bibfnamefont {H.}~\bibnamefont {Zbinden}}, \emph {et~al.},\ }\bibfield  {title} {\bibinfo {title} {High-efficiency superconducting nanowire single-photon detectors fabricated from mosi thin-films},\ }\href@noop {} {\bibfield  {journal} {\bibinfo  {journal} {Optics express}\ }\textbf {\bibinfo {volume} {23}},\ \bibinfo {pages} {33792} (\bibinfo {year} {2015})}\BibitemShut {NoStop}%
\bibitem [{\citenamefont {Gourgues}\ \emph {et~al.}(2019)\citenamefont {Gourgues}, \citenamefont {Los}, \citenamefont {Zichi}, \citenamefont {Chang}, \citenamefont {Kalhor}, \citenamefont {Bulgarini}, \citenamefont {Dorenbos}, \citenamefont {Zwiller},\ and\ \citenamefont {Zadeh}}]{gourgues2019superconducting}%
  \BibitemOpen
  \bibfield  {author} {\bibinfo {author} {\bibfnamefont {R.}~\bibnamefont {Gourgues}}, \bibinfo {author} {\bibfnamefont {J.~W.}\ \bibnamefont {Los}}, \bibinfo {author} {\bibfnamefont {J.}~\bibnamefont {Zichi}}, \bibinfo {author} {\bibfnamefont {J.}~\bibnamefont {Chang}}, \bibinfo {author} {\bibfnamefont {N.}~\bibnamefont {Kalhor}}, \bibinfo {author} {\bibfnamefont {G.}~\bibnamefont {Bulgarini}}, \bibinfo {author} {\bibfnamefont {S.~N.}\ \bibnamefont {Dorenbos}}, \bibinfo {author} {\bibfnamefont {V.}~\bibnamefont {Zwiller}},\ and\ \bibinfo {author} {\bibfnamefont {I.~E.}\ \bibnamefont {Zadeh}},\ }\bibfield  {title} {\bibinfo {title} {Superconducting nanowire single photon detectors operating at temperature from 4 to 7 k},\ }\href@noop {} {\bibfield  {journal} {\bibinfo  {journal} {Optics express}\ }\textbf {\bibinfo {volume} {27}},\ \bibinfo {pages} {24601} (\bibinfo {year} {2019})}\BibitemShut {NoStop}%
\bibitem [{\citenamefont {Esmaeil~Zadeh}\ \emph {et~al.}(2017)\citenamefont {Esmaeil~Zadeh}, \citenamefont {Los}, \citenamefont {Gourgues}, \citenamefont {Steinmetz}, \citenamefont {Bulgarini}, \citenamefont {Dobrovolskiy}, \citenamefont {Zwiller},\ and\ \citenamefont {Dorenbos}}]{esmaeil2017single}%
  \BibitemOpen
  \bibfield  {author} {\bibinfo {author} {\bibfnamefont {I.}~\bibnamefont {Esmaeil~Zadeh}}, \bibinfo {author} {\bibfnamefont {J.~W.}\ \bibnamefont {Los}}, \bibinfo {author} {\bibfnamefont {R.}~\bibnamefont {Gourgues}}, \bibinfo {author} {\bibfnamefont {V.}~\bibnamefont {Steinmetz}}, \bibinfo {author} {\bibfnamefont {G.}~\bibnamefont {Bulgarini}}, \bibinfo {author} {\bibfnamefont {S.~M.}\ \bibnamefont {Dobrovolskiy}}, \bibinfo {author} {\bibfnamefont {V.}~\bibnamefont {Zwiller}},\ and\ \bibinfo {author} {\bibfnamefont {S.~N.}\ \bibnamefont {Dorenbos}},\ }\bibfield  {title} {\bibinfo {title} {Single-photon detectors combining high efficiency, high detection rates, and ultra-high timing resolution},\ }\href@noop {} {\bibfield  {journal} {\bibinfo  {journal} {Apl Photonics}\ }\textbf {\bibinfo {volume} {2}} (\bibinfo {year} {2017})}\BibitemShut {NoStop}%
\bibitem [{\citenamefont {Oton}(2015)}]{oton2015long}%
  \BibitemOpen
  \bibfield  {author} {\bibinfo {author} {\bibfnamefont {C.}~\bibnamefont {Oton}},\ }\bibfield  {title} {\bibinfo {title} {Long-working-distance grating coupler for integrated optical devices},\ }\href@noop {} {\bibfield  {journal} {\bibinfo  {journal} {IEEE Photonics Journal}\ }\textbf {\bibinfo {volume} {8}},\ \bibinfo {pages} {1} (\bibinfo {year} {2015})}\BibitemShut {NoStop}%
\bibitem [{\citenamefont {Khan}\ \emph {et~al.}(2020)\citenamefont {Khan}, \citenamefont {Combri{\'e}},\ and\ \citenamefont {De~Rossi}}]{khan2020long}%
  \BibitemOpen
  \bibfield  {author} {\bibinfo {author} {\bibfnamefont {M.~S.~I.}\ \bibnamefont {Khan}}, \bibinfo {author} {\bibfnamefont {S.}~\bibnamefont {Combri{\'e}}},\ and\ \bibinfo {author} {\bibfnamefont {A.}~\bibnamefont {De~Rossi}},\ }\bibfield  {title} {\bibinfo {title} {Long working distance apodized grating coupler},\ }\href@noop {} {\bibfield  {journal} {\bibinfo  {journal} {arXiv preprint arXiv:2006.16247}\ } (\bibinfo {year} {2020})}\BibitemShut {NoStop}%
\bibitem [{\citenamefont {Raussendorf}\ \emph {et~al.}(2003)\citenamefont {Raussendorf}, \citenamefont {Browne},\ and\ \citenamefont {Briegel}}]{raussendorf2003measurement}%
  \BibitemOpen
  \bibfield  {author} {\bibinfo {author} {\bibfnamefont {R.}~\bibnamefont {Raussendorf}}, \bibinfo {author} {\bibfnamefont {D.~E.}\ \bibnamefont {Browne}},\ and\ \bibinfo {author} {\bibfnamefont {H.~J.}\ \bibnamefont {Briegel}},\ }\bibfield  {title} {\bibinfo {title} {Measurement-based quantum computation on cluster states},\ }\href@noop {} {\bibfield  {journal} {\bibinfo  {journal} {Physical review A}\ }\textbf {\bibinfo {volume} {68}},\ \bibinfo {pages} {022312} (\bibinfo {year} {2003})}\BibitemShut {NoStop}%
\bibitem [{\citenamefont {Briegel}\ \emph {et~al.}(2009)\citenamefont {Briegel}, \citenamefont {Browne}, \citenamefont {D{\"u}r}, \citenamefont {Raussendorf},\ and\ \citenamefont {Van~den Nest}}]{briegel2009measurement}%
  \BibitemOpen
  \bibfield  {author} {\bibinfo {author} {\bibfnamefont {H.~J.}\ \bibnamefont {Briegel}}, \bibinfo {author} {\bibfnamefont {D.~E.}\ \bibnamefont {Browne}}, \bibinfo {author} {\bibfnamefont {W.}~\bibnamefont {D{\"u}r}}, \bibinfo {author} {\bibfnamefont {R.}~\bibnamefont {Raussendorf}},\ and\ \bibinfo {author} {\bibfnamefont {M.}~\bibnamefont {Van~den Nest}},\ }\bibfield  {title} {\bibinfo {title} {Measurement-based quantum computation},\ }\href@noop {} {\bibfield  {journal} {\bibinfo  {journal} {Nature Physics}\ }\textbf {\bibinfo {volume} {5}},\ \bibinfo {pages} {19} (\bibinfo {year} {2009})}\BibitemShut {NoStop}%
\bibitem [{\citenamefont {Charaev}\ \emph {et~al.}(2023)\citenamefont {Charaev}, \citenamefont {Bandurin}, \citenamefont {Bollinger}, \citenamefont {Phinney}, \citenamefont {Drozdov}, \citenamefont {Colangelo}, \citenamefont {Butters}, \citenamefont {Taniguchi}, \citenamefont {Watanabe}, \citenamefont {He} \emph {et~al.}}]{charaev2023single}%
  \BibitemOpen
  \bibfield  {author} {\bibinfo {author} {\bibfnamefont {I.}~\bibnamefont {Charaev}}, \bibinfo {author} {\bibfnamefont {D.}~\bibnamefont {Bandurin}}, \bibinfo {author} {\bibfnamefont {A.}~\bibnamefont {Bollinger}}, \bibinfo {author} {\bibfnamefont {I.}~\bibnamefont {Phinney}}, \bibinfo {author} {\bibfnamefont {I.}~\bibnamefont {Drozdov}}, \bibinfo {author} {\bibfnamefont {M.}~\bibnamefont {Colangelo}}, \bibinfo {author} {\bibfnamefont {B.}~\bibnamefont {Butters}}, \bibinfo {author} {\bibfnamefont {T.}~\bibnamefont {Taniguchi}}, \bibinfo {author} {\bibfnamefont {K.}~\bibnamefont {Watanabe}}, \bibinfo {author} {\bibfnamefont {X.}~\bibnamefont {He}}, \emph {et~al.},\ }\bibfield  {title} {\bibinfo {title} {Single-photon detection using high-temperature superconductors},\ }\href@noop {} {\bibfield  {journal} {\bibinfo  {journal} {Nature Nanotechnology}\ }\textbf {\bibinfo {volume} {18}},\ \bibinfo {pages} {343} (\bibinfo {year} {2023})}\BibitemShut {NoStop}%
\bibitem [{\citenamefont {Ziabari}\ \emph {et~al.}(2010)\citenamefont {Ziabari}, \citenamefont {Bian},\ and\ \citenamefont {Shakouri}}]{ziabari2010adaptive}%
  \BibitemOpen
  \bibfield  {author} {\bibinfo {author} {\bibfnamefont {A.}~\bibnamefont {Ziabari}}, \bibinfo {author} {\bibfnamefont {Z.}~\bibnamefont {Bian}},\ and\ \bibinfo {author} {\bibfnamefont {A.}~\bibnamefont {Shakouri}},\ }\bibfield  {title} {\bibinfo {title} {Adaptive power blurring techniques to calculate ic temperature profile under large temperature variations},\ }\href@noop {} {\bibfield  {journal} {\bibinfo  {journal} {International Microelectronic and Packaging Society (IMAPS) ATW on Thermal Management}\ ,\ \bibinfo {pages} {28}} (\bibinfo {year} {2010})}\BibitemShut {NoStop}%
\bibitem [{\citenamefont {Pernice}\ \emph {et~al.}(2012)\citenamefont {Pernice}, \citenamefont {Schuck}, \citenamefont {Minaeva}, \citenamefont {Li}, \citenamefont {Goltsman}, \citenamefont {Sergienko},\ and\ \citenamefont {Tang}}]{pernice2012high}%
  \BibitemOpen
  \bibfield  {author} {\bibinfo {author} {\bibfnamefont {W.~H.}\ \bibnamefont {Pernice}}, \bibinfo {author} {\bibfnamefont {C.}~\bibnamefont {Schuck}}, \bibinfo {author} {\bibfnamefont {O.}~\bibnamefont {Minaeva}}, \bibinfo {author} {\bibfnamefont {M.}~\bibnamefont {Li}}, \bibinfo {author} {\bibfnamefont {G.}~\bibnamefont {Goltsman}}, \bibinfo {author} {\bibfnamefont {A.}~\bibnamefont {Sergienko}},\ and\ \bibinfo {author} {\bibfnamefont {H.}~\bibnamefont {Tang}},\ }\bibfield  {title} {\bibinfo {title} {High-speed and high-efficiency travelling wave single-photon detectors embedded in nanophotonic circuits},\ }\href@noop {} {\bibfield  {journal} {\bibinfo  {journal} {Nature communications}\ }\textbf {\bibinfo {volume} {3}},\ \bibinfo {pages} {1325} (\bibinfo {year} {2012})}\BibitemShut {NoStop}%
\bibitem [{\citenamefont {Amirfeiz}\ \emph {et~al.}(2000)\citenamefont {Amirfeiz}, \citenamefont {Bengtsson}, \citenamefont {Bergh}, \citenamefont {Zanghellini},\ and\ \citenamefont {B{\"o}rjesson}}]{amirfeiz2000formation}%
  \BibitemOpen
  \bibfield  {author} {\bibinfo {author} {\bibfnamefont {P.}~\bibnamefont {Amirfeiz}}, \bibinfo {author} {\bibfnamefont {S.}~\bibnamefont {Bengtsson}}, \bibinfo {author} {\bibfnamefont {M.}~\bibnamefont {Bergh}}, \bibinfo {author} {\bibfnamefont {E.}~\bibnamefont {Zanghellini}},\ and\ \bibinfo {author} {\bibfnamefont {L.}~\bibnamefont {B{\"o}rjesson}},\ }\bibfield  {title} {\bibinfo {title} {Formation of silicon structures by plasma-activated wafer bonding},\ }\href@noop {} {\bibfield  {journal} {\bibinfo  {journal} {Journal of the Electrochemical Society}\ }\textbf {\bibinfo {volume} {147}},\ \bibinfo {pages} {2693} (\bibinfo {year} {2000})}\BibitemShut {NoStop}%
\bibitem [{\citenamefont {Resta}\ \emph {et~al.}(2023)\citenamefont {Resta}, \citenamefont {Stasi}, \citenamefont {Perrenoud}, \citenamefont {El-Khoury}, \citenamefont {Brydges}, \citenamefont {Thew}, \citenamefont {Zbinden},\ and\ \citenamefont {Bussi{\`e}res}}]{resta2023gigahertz}%
  \BibitemOpen
  \bibfield  {author} {\bibinfo {author} {\bibfnamefont {G.~V.}\ \bibnamefont {Resta}}, \bibinfo {author} {\bibfnamefont {L.}~\bibnamefont {Stasi}}, \bibinfo {author} {\bibfnamefont {M.}~\bibnamefont {Perrenoud}}, \bibinfo {author} {\bibfnamefont {S.}~\bibnamefont {El-Khoury}}, \bibinfo {author} {\bibfnamefont {T.}~\bibnamefont {Brydges}}, \bibinfo {author} {\bibfnamefont {R.}~\bibnamefont {Thew}}, \bibinfo {author} {\bibfnamefont {H.}~\bibnamefont {Zbinden}},\ and\ \bibinfo {author} {\bibfnamefont {F.}~\bibnamefont {Bussi{\`e}res}},\ }\bibfield  {title} {\bibinfo {title} {Gigahertz detection rates and dynamic photon-number resolution with superconducting nanowire arrays},\ }\href@noop {} {\bibfield  {journal} {\bibinfo  {journal} {Nano Letters}\ } (\bibinfo {year} {2023})}\BibitemShut {NoStop}%
\bibitem [{Dar(2023)}]{Darpa_INSPIRED}%
  \BibitemOpen
  \href@noop {} {\bibinfo {title} {Broad agency announcement intensity-squeezed photonic integration for revolutionary detectors (inspired) microsystems technology office hr001123s0052}},\ \bibinfo {howpublished} {\url{https://www.darpa.mil/attachments/HR001123S0052-Amendment-01.pdf}} (\bibinfo {year} {2023}),\ \bibinfo {note} {accessed: 2023-11-15}\BibitemShut {NoStop}%
\end{thebibliography}%

\end{document}